\documentclass[a4paper,11pt]{article}
\usepackage{jinstpub} 
\usepackage{lineno}
\usepackage{subcaption}
\usepackage{upgreek}
\usepackage{siunitx}
\usepackage{tabularx}
\usepackage{hyperref}
\DeclareSIUnit{\SLPM}{SLPM}
\DeclareSIUnit{\ppm}{ppm}
\title{\boldmath MOTION, a liquid xenon time projection chamber platform for high voltage technologies in dark matter detectors}

\author[*]{Y.~Biondi\note[*]{Corresponding author.},}
\author{A.~Jansen,}
\author{K.~Ding,}
\author{T.~Sonius,}
\author{M.~Schrank,}
\author{A.~Schwenck}
\affiliation{Institute for Astroparticle Physics, Karlsruhe Institute of Technology, Hermann-von-Helmholtz-Platz 1, 76344 Eggenstein-Leopoldshafen, Germany}

\emailAdd{yanina.biondi@kit.edu}

\abstract{The XLZD observatory is a next-generation experiment designed to search for weakly interacting massive particles (WIMPs) and other rare events using a \SIrange{60}{80}{tonne} liquid xenon time projection chamber (TPC). This detector aims to achieve sensitivity across the full WIMP parameter space down to the neutrino fog, establishing the ultimate sensitivity for this dark matter search paradigm. This unprecedented scale introduces substantial engineering challenges and pushes operation into largely unexplored regimes: the interplay between high-voltage (HV) systems, liquid xenon, and conducting materials in ultra-pure environments. To systematically investigate these challenges, we have built MOTION, a \SI{70}{kg} LXe detector dedicated to understanding HV performance and electrostatic phenomena up to \SI{200}{kV} (negative polarity). We describe the design and construction of the experimental infrastructure, including the cryogenic system, xenon purification and storage. MOTION enables controlled studies of dielectric breakdown in LXe, permitting systematic characterization of discharge mechanisms and their dependence on electrode geometry, surface condition, and applied voltage. The detector also facilitates investigations of field emission and photoemission from electrodes following various surface treatments, and provides a platform for validating the design of an HV feedthrough constructed from radiopure materials. The insights from these studies are essential for ensuring the operational stability, radiopurity, and scalability required for next-generation dark matter detectors.}


\arxivnumber{xxxx.xxxxx} 

\begin{document}
\maketitle
\flushbottom

\section{Introduction}

One of the most fundamental open questions in physics is that of the nature and composition of dark matter. One particularly promising direct detection method is the use of dual-phase liquid xenon (LXe) time-projection chambers (TPCs) to search for interactions between Weakly Interacting Massive Particles (WIMPs) and the target xenon. More than a decade of effort by several experiments has led to the most stringent exclusion bounds for WIMP masses above a few GeV/c$^{2}$~\cite{XENON:2025vwd,LZ:2024zvo,PandaX:2024qfu,XENON:2018voc,LUX:2016ggv}. These detectors have been successfully scaled from a few kilograms to the tonne scale, while simultaneously suppressing the background count rate per kilogram of detector material by a factor of 30,000 over about 15 years~\cite{Baudis:2025yva}. 

The XLZD Observatory is a next-generation liquid-xenon experiment. Its baseline design includes \SI{60}{tonnes} of LXe as the active target in a cylindrical TPC with dimensions of \SI{3}{m} in diameter and height, a two-fold increase in spatial dimensions compared to current running detectors~\cite{XLZD:2024nsu}. Together with an ultra-low background level, this large target mass makes the detector a versatile observatory for rare-event processes in nuclear, particle and astroparticle physics. While the search for WIMP dark matter is the prime scientific goal, other important channels open up~\cite{Aalbers:2022dzr}, including a competitive search for neutrinoless double beta decay employing natural xenon without enrichment~\cite{XLZD:2024pdv}.

The drift and extraction fields are key components in the success of LXe TPCs, as position reconstruction, particle identification, and energy reconstruction depend strongly on them. Previous and current LXe TPCs have faced difficulties biasing their cathode to HV, in the order of tens of kV (negative polarity), and none has reached its design value for the drift field~\cite{XENON:2024wpa,XENON:2020kmp,LZ:2019sgr,LZ:2024zvo}. Analogously, biasing the anode and ground electrode, which defines the extraction field in the range of \SIrange{2}{10}{kV/cm}, has also been challenging. These difficulties are believed to arise from diverse phenomena spanning different fields that are not fully understood. 
Regions subjected to a high electric field are more likely to produce secondary electron emission. The exact mechanism for this emission remains unknown, and it has been proposed that the creation of an ion monolayer on the surface of the conductor material~\cite{Malter1936}, impurities trapped in the gas–liquid interface~\cite{Kopec2021}, cathodic emission~\cite{Tomas:2018pny}, fluorescence in materials or the photoelectric effect on conductors might be responsible. This secondary electron emission can produce electron trains detected in the photosensors that do not match any interaction in the TPC, and thus introduce additional background~\cite{Kopec2021}, affecting the detector performance. Successful voltage delivery to the electrodes and stable performance of the drift and extraction field is therefore entwined with the full characterisation of the HV components, understanding the power supply, careful design of the HV feedthrough, the connection to the electrodes, and the electrodes themselves.

Although the predicted bulk breakdown strength of LXe is near \SI{1}{MV/cm}, experiments consistently observe breakdown at much lower fields, as low as \SI{-50}{kV/cm}~\cite{Xebra}. Additionally, discharges are not a deterministic event, and more information about the probability distribution around different breakdown voltages needs to be gathered to characterise HV components and their survival probability based on the applied voltage~\cite{kuffel2000high}. Past measurements suggest that surface effects, rather than bulk ionization, dominate~\cite{breakdown, GasBreakdown}. In particular, local field enhancement at electrode asperities and the dependence of breakdown field on stressed electrode area (SEA) have been identified as key factors~\cite{Xebra}. The SEA of next-generation experiments will grow with their linear dimensions, making it necessary to develop a program to identify faults and to characterise and test HV components.

The MOTION (experiMent for develOpmenT of technologies in liquId xenON) detector is a \SI{70}{kg} LXe detector that provides a unique platform for systematic studies of the development of reliable HV components, dielectric breakdown and discharge phenomena in LXe. 

The paper is organised as follows. Section 2 gives an overview of the facility infrastructure, and Section 3 the choices behind the pressure and vacuum vessels and the gas system. Section 4 describes the instrumentation inside the inner vessel, where the detector is located, and Section 5 the monitoring, alarm, database and visualisation systems. Section 6 covers the commissioning of the facility through the injection, condensation and recuperation of the first xenon in the detector. Section 7 presents the protocol developed to characterise conductor surfaces by laser microscopy, Section 8 outlines the planned upgrades and measurements, and Section 9 summarises the work.

\label{sec:intro}

\section{Facility infrastructure}
\begin{figure}[h!]
    \centering
    \includegraphics[width=\textwidth]{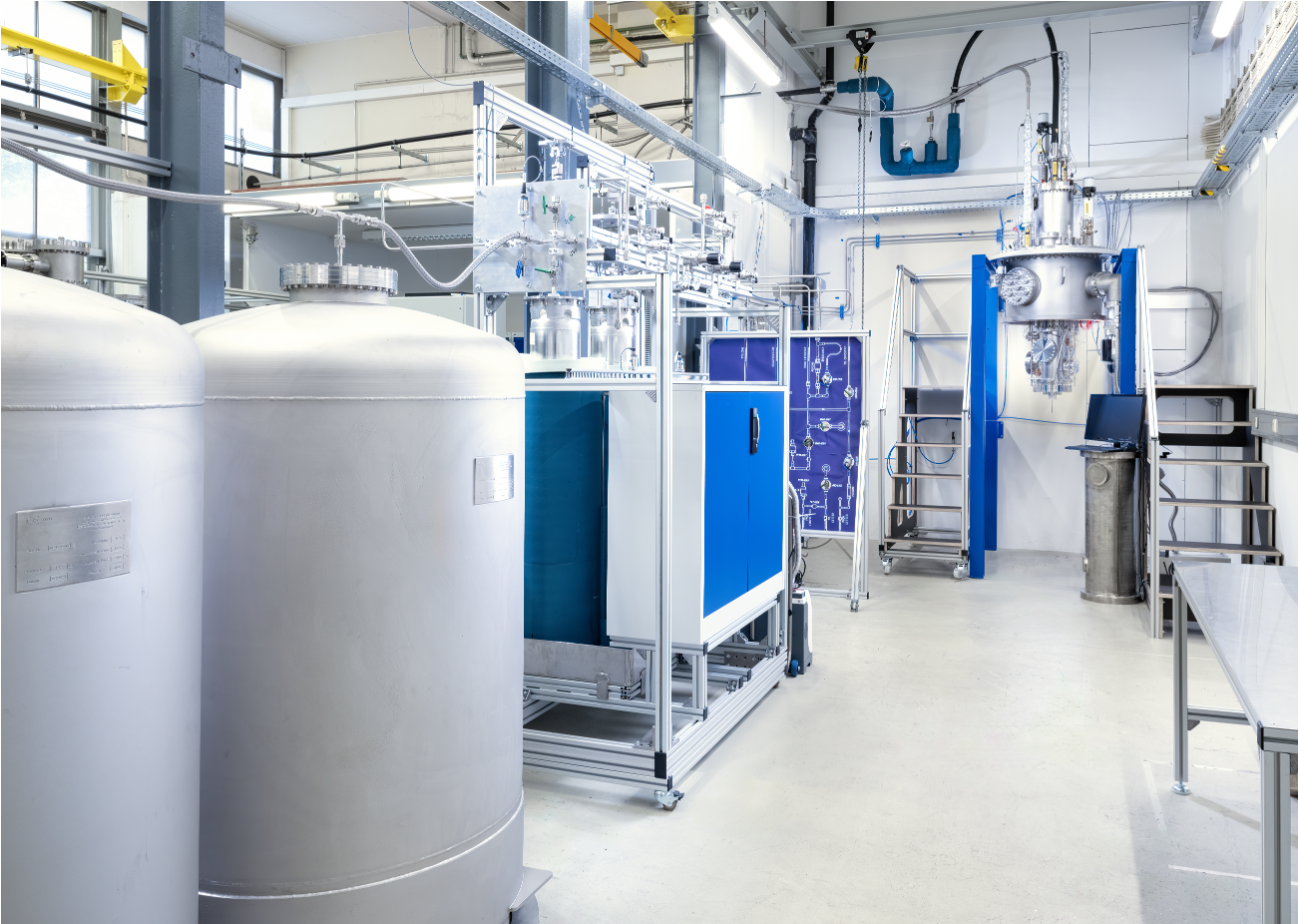}
    \caption{Xenon laboratory containing the emergency vessels, electric cabinet, xenon gas bottles, gas purification panels, support structure and MOTION cryostat. Credit: A. Bramsiepe / KIT}
    \label{fig:lab_xenon}
\end{figure}

Figure~\ref{fig:lab_xenon} shows the layout of the xenon laboratory located at the Karlsruhe Institute of Technology (KIT), Institute for Astroparticle Physics.

\begin{itemize}
    \item \textbf{Xenon storage} consists of two custom-made, identical stainless steel pressure vessels, suitable for low-temperature operation, each storing up to \SI{35}{kg} of xenon.
    \item \textbf{High-pressure panel} controls the gas flow from and to the xenon storage vessels.
    \item \textbf{Gas purification panel} to recirculate the xenon in the cryostat through a closed loop, where fine and ultra-fine filters, along with a getter, remove particulates and electronegative impurities diffused in the xenon. The xenon gas is recirculated by a compressor.
    \item \textbf{Liquid nitrogen (LN$_2$) supply and condenser} comprise a \SI{20000}{\litre} LN$_2$ storage tank located outside the laboratory building, connected to the condenser via a cryogenic transfer line. The condenser incorporates a cold head in thermal contact with an LN$_2$ bath, providing the cooling capacity required to liquefy xenon.
 
    \item \textbf{Cryostat} consists of an outer vessel kept at high vacuum to minimise heat transfer, an inner vessel operated at 1.8 bar, containing liquid xenon beneath a gaseous xenon phase. The cryostat has ports in all directions to accommodate HV feedthroughs from the sides and bottom of the chamber, as well as other detector expansion modules, such as LXe purity monitors.
    \item \textbf{Emergency vessels} that can contain the gaseous xenon in the vessels after a rupture of safety burst disks. The emergency vessels can also help lower the pressure inside the inner vessel in the event of an unexpected pressure increase due to a loss of cooling power.
    \item \textbf{Support structure and crane} to handle the expected weight of approximately \SI{1.3}{tonnes} of the entire system during assembly and maintenance.
    \item \textbf{Slow control system} continuously monitors pressure, xenon gas flow, and temperatures in the system. Safety interlocks and SMS alarms respond to any unexpected behaviour, to ensure the integrity of the experiment and the safety of its operators. The values are stored in a database for later retrieval.
    \end{itemize}

The following section describes more in detail each one of these subsystems.
\section{Gas handling and cryogenics}

\begin{figure}[h!]
    \centering
    \includegraphics[width=0.8\textwidth]{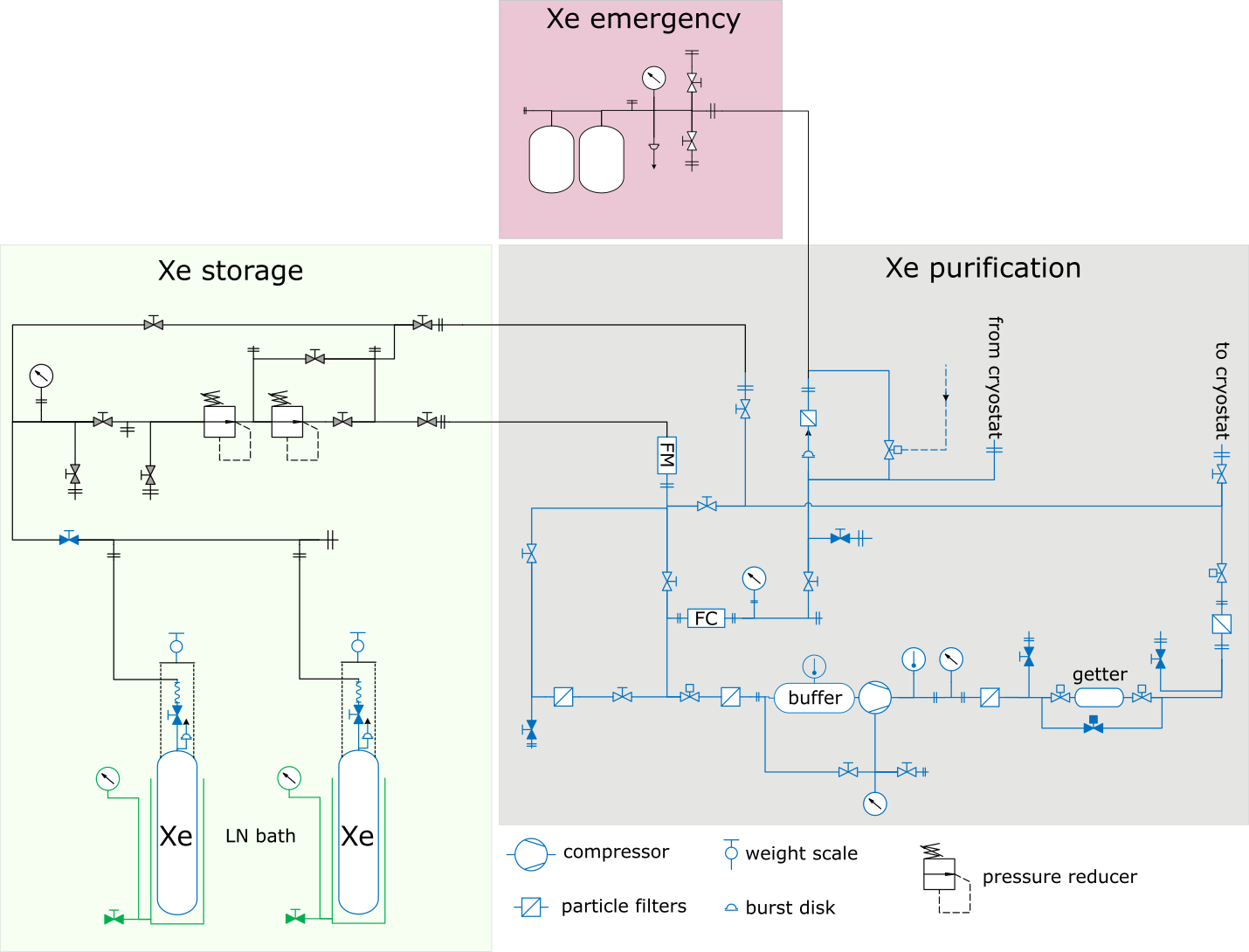}
    \caption{P\&ID of the gas system, installed in the laboratory. The xenon storage system handles the higher pressure required to access xenon inside the stainless steel bottles and contains two pressure reducers that bring the pressure down to the operational pressure, around \SI{1.8}{bar}. The xenon purification panel recirculates xenon extracted from the detector using a compressor. The xenon passes through a series of filters and a getter, while its flow is controlled and measured using flow controllers and flow meters, respectively. The xenon emergency panel contains two buffer vessels that transfer xenon from the detector, via hydraulic bypass valves or burst disks, to provide a pressure relief in the event of faulty user operation or cooling power failure.}
    \label{fig:motion_PID}
\end{figure}

All components manufactured for the system comply with the Pressure Equipment Directive 2014/68/EU (PED), with the technical rules of AD 2000 applied. This includes the storage bottles, the emergency vessels, and the cryostat vessels. 
Figure~\ref{fig:motion_PID} shows the Piping and instrumentation diagram (P\&ID) of the xenon storage, the purification panel, and the emergency buffer vessels. 

\subsection{Storage system and high-pressure panel}
 Two identical custom-made stainless steel pressure vessels store all the xenon used during operation, up to \SI{70}{kg}. The safety device, a burst disc, is directly attached to each vessel, which protects the vessels from bursting during a fire or overfilling. Both vessels sit inside open stainless steel cylinders that can be filled with liquid nitrogen (LN$_2$) from a mobile dewar to cool the vessels, allowing xenon from the cryostat inner vessel, the tubing, or the emergency vessels to be cryo-pumped back into storage. With a vapour pressure of approximately \SI{60}{bar} under standard conditions, the pressure inside both vessels remains well below the burst disc's trigger pressure and, hence, well below the vessels' design pressure~\cite{NIST}. This is valid for a filling of up to \SI{35}{kg} of xenon each. A scale that continuously measures vessel weight, instrumented with a load cell (LC SLS410, Mettler Toledo) and a transducer (Omega TXDIN1600S), is installed to prevent overfilling and monitor the xenon mass.
 
The xenon storage panel is a welded tube-and-valve assembly that manually controls gas flow to and from the xenon storage vessels. The system is designed so that a standard \SI{200}{bar} gas bottle can be attached to the high-pressure side. Each device can withstand the maximum pressure of the storage vessels. The panel controls the flow of xenon to the inner vessel using two pressure reducers, and a bypass allows flow from the inner vessel back to the xenon storage bottles.

\subsection{Gas purification panel}
The purification panel is an assembly of tubes, manual valves, automatic valves, two flow meters, a flow controller, a circulation pump, a getter, and filters. The purpose of the panel is to clean the xenon continuously during operation with LXe inside the detector. The extracted liquid is evaporated by the time it reaches the purification panel due to thermal conduction in the tubing, and a flow meter controller (Bronkhorst F-201CM) sets the recirculation speed before it passes by a series of filters (SS-4TF-TW and SS-SCF3-VR4-P-30 from Swagelok), reaches the compressor (KNF N630ST.12E), and gets purified by the getter (SAES Monotorr PS4-MT3-N-2). All welded connections have been pressure tested prior to installation to a value well above the operating pressure. The compressor membrane is susceptible to damage when pumping xenon owing to its high atomic mass. Such failures are detected by monitoring the pressure in the inner space of the pump, between the working and the safety diaphragm.

\subsection{Emergency storage}
The emergency storage consists of two vessels with \SI{1100}{\litre} volume each. The vessels are interconnected and are connected to the detector's inner vessel via a burst disc. If the pressure inside the inner vessel exceeds the burst pressure in the case of failure, the xenon will be released into the emergency vessels. The emergency vessels themselves are further protected by a burst disc. The overall system is designed so that all xenon stored at its vapour pressure inside the storage vessels (two times \SI{35}{kg}) can expand into the whole system, resulting in a gas pressure of less than \SI{8}{bar}. In such a case, the xenon can be recuperated from the emergency panels by cryo-pumping it back into the storage vessels and then addressing the cause of failure.

\subsection{Cryostat}


The cryostat consists of two nested vessels: the outer and the inner. The outer vessel is a common insulation vacuum container with a volume of approximately \SI{600}{L}. Inside the outer vessel is the inner vessel, which contains xenon. As the xenon tests involve both cryogenics and HV, the vessel is equipped with feedthroughs in all directions, such as gas inlets and outlets, HV feedthroughs, and electrical feedthroughs.
The outer vessel is properly connected to ground potential and is also equipped with a drop-out plate in case of leakage from the inner parts, ensuring a maximum operating pressure well below \SI{0.5}{bar}.

The cryostat inner vessel is the actual LXe container, in which new technologies for future LXe experiments will be tested. It is installed inside the outer vessel and connected to the outside setup only via electrical feedthroughs and flexible DN10 tubing with 1/2" VCR connectors. Including the xenon collector, which is attached to the condenser, the overall volume is approximately \SI{28}{\litre}. The pressure in the inner vessel is overseen by a pressure transducer (Endress+Hauser PMC21). Another pressure transducer is connected to a thin piping system that measures the pressure at the bottom of the detector, providing an independent method of measuring the liquid level inside the inner vessel.

The total static heat load on the inner vessel is about \SI{12}{W}: \SI{1.4}{W} from the support structure (seven M12 rods), \SI{5.5}{W} from cabling, and \SI{5}{W} from thermal radiation. The total heat transfer from the gas flexible tubing can be neglected. 

\subsection{Condenser}

The xenon liquefaction system uses LN$_2$ to condense gaseous xenon (GXe) at its saturation temperature while avoiding the risk of freezing. The condenser design was based on an already tested design at the University of Freiburg~\cite{pancake}. A reservoir of LN$_2$ at the top of the assembly provides a constant-temperature cold source at \SI{77}{K}, maintained by boiling nitrogen at constant pressure. Because this temperature lies well below the freezing point of xenon (\SI{161}{K}), the LN$_2$ is not coupled directly to the xenon; instead, a sealed, constant-volume gaseous-nitrogen (GN$_2$) buffer volume is placed between two heat exchangers and serves as a passive thermal link between the reservoir and the condenser. Within this buffer, heat is transported upward by natural convection: gas warmed at the lower heat exchanger rises, is cooled against the LN$_2$ reservoir at the upper heat exchanger, and sinks again, establishing a self-sustaining thermosiphon. Since the volume is fixed, its pressure and density set the effective thermal conductance between the two heat exchangers, so that the temperature of the lower heat exchanger can be held near the xenon saturation point (approximately \SI{174}{K}) rather than at \SI{77}{K}. Warm GXe entering below this heat exchanger is thereby cooled and condensed, and the resulting LXe drains into the inner vessel, which holds the xenon in gas–liquid equilibrium. Heat leaking into the vessel by radiation continuously boils off a small fraction of the liquid; this vapour leaves as GXe exhaust and is recirculated to the condenser inlet, closing the xenon loop. A separate LN$_2$ line at the base of the vessel provides pre-cooling during the initial cooldown and offsets part of the steady-state heat load. Simulations and testing were used to estimate the total cooling power around \SI{130}{W}.

A \SI{2.25}{\litre} buffer vessel contains GN$_2$ and is certified for pressures up to \SI{100}{\bar}. It supplies GN2 to the (\SI{0.4}{\litre}) nitrogen heat exchanger located between the LN$_2$ supply and the inner vessel, and can itself be refilled from the lab's nitrogen supply. Since nitrogen is cooled and partially liquefied inside the heat exchanger, and the gas volume can be trapped, an additional safety valve is installed as close to the heat exchanger as possible to ensure safe operation. 

\begin{figure}[h!]
    \centering
    \begin{subfigure}[b]{0.50\textwidth}
         \centering
         \includegraphics[width=\textwidth]{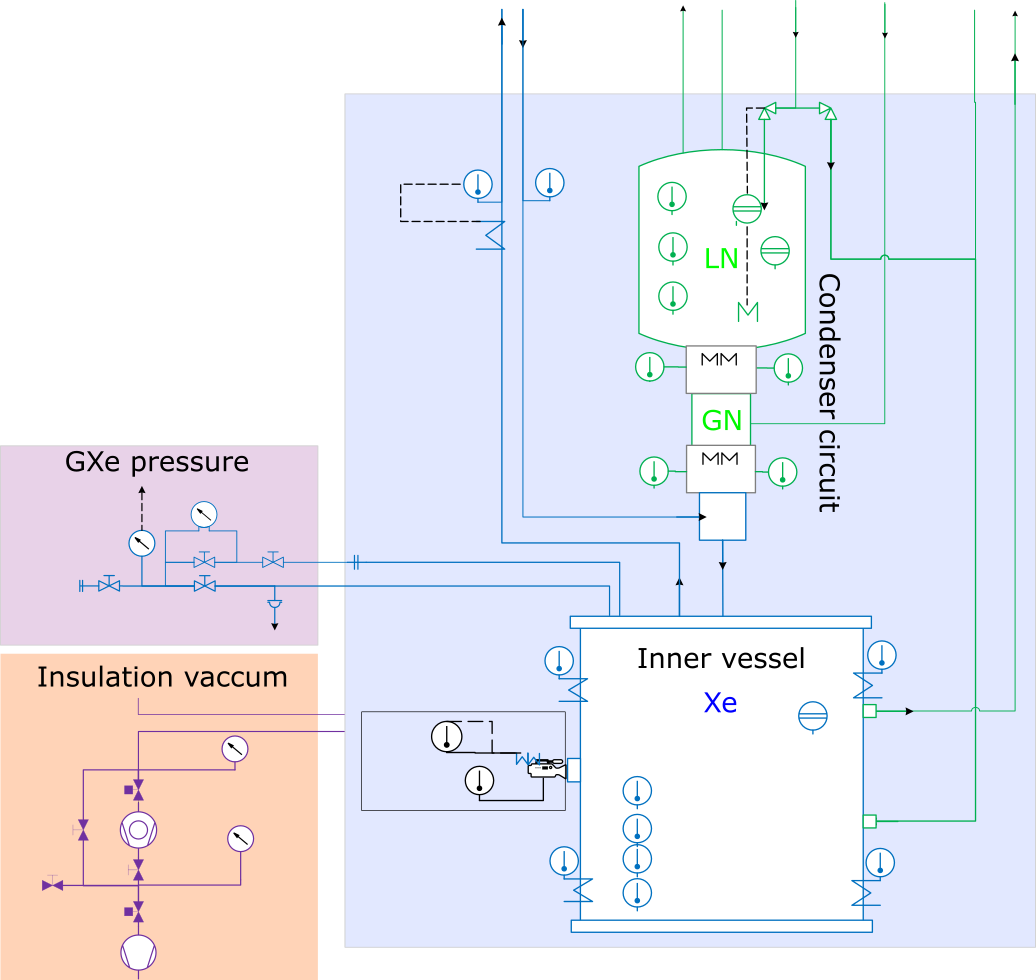}
    \end{subfigure}
    \begin{subfigure}[b]{0.43\textwidth}
         \centering
         \includegraphics[width=\textwidth]{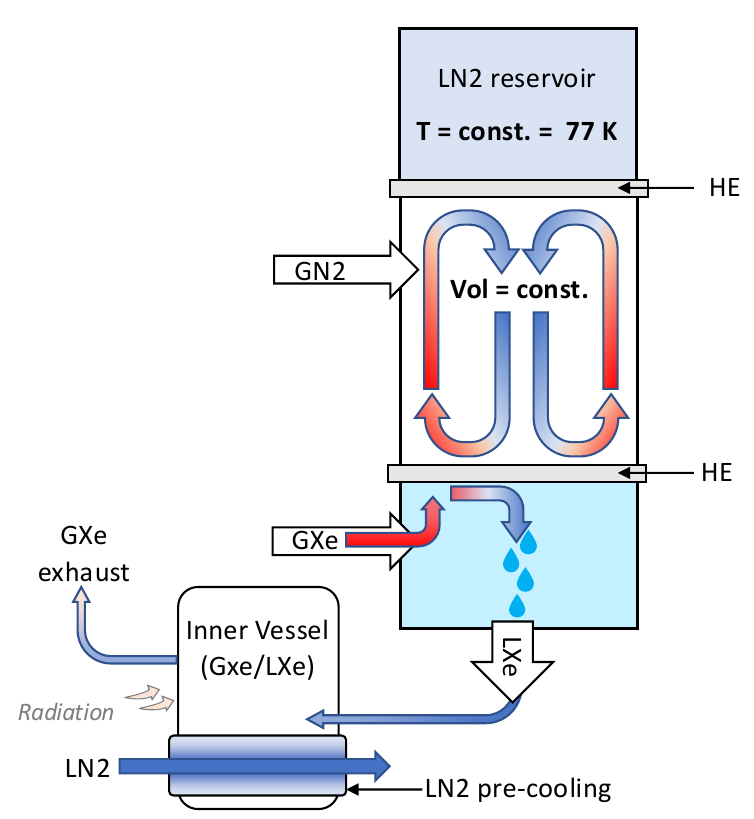}
    \end{subfigure}
\caption{Left: P\&ID of the cryostat and condenser. Right: Schematic of the liquid-nitrogen–driven xenon liquefaction system. Liquid nitrogen in the upper reservoir provides a constant-temperature cold source at \SI{77}{K}. A sealed, constant-volume GN$_2$ buffer between the two heat exchangers (HE) transports heat upward by natural convection (thermosiphon; red/blue arrows indicate warm rising and cooled falling gas) and acts as a passive thermal regulator, allowing the lower HE to sit near the xenon saturation point. Incoming warm GXe is cooled at the lower HE and condenses into LXe, which drains into the inner vessel holding the GXe/LXe mixture. Heat leaking in by radiation or high recirculation speed boils off xenon that leaves as GXe exhaust and is recirculated, while a separate LN$_2$ line provides pre-cooling of the inner vessel.}
\label{fig:cryogenics_PID}
\end{figure}

Figure~\ref{fig:cryogenics_PID} shows the P\&ID of the cryostat and condenser and its working principle, which provides the cooling power necessary to cool and liquefy xenon gas at different recirculation flow speeds.

\section{Detector and instrumentation}

\begin{figure}[!htbp]
    \centering
    \begin{subfigure}[b]{0.42\textwidth}
         \centering
         \includegraphics[width=\textwidth]{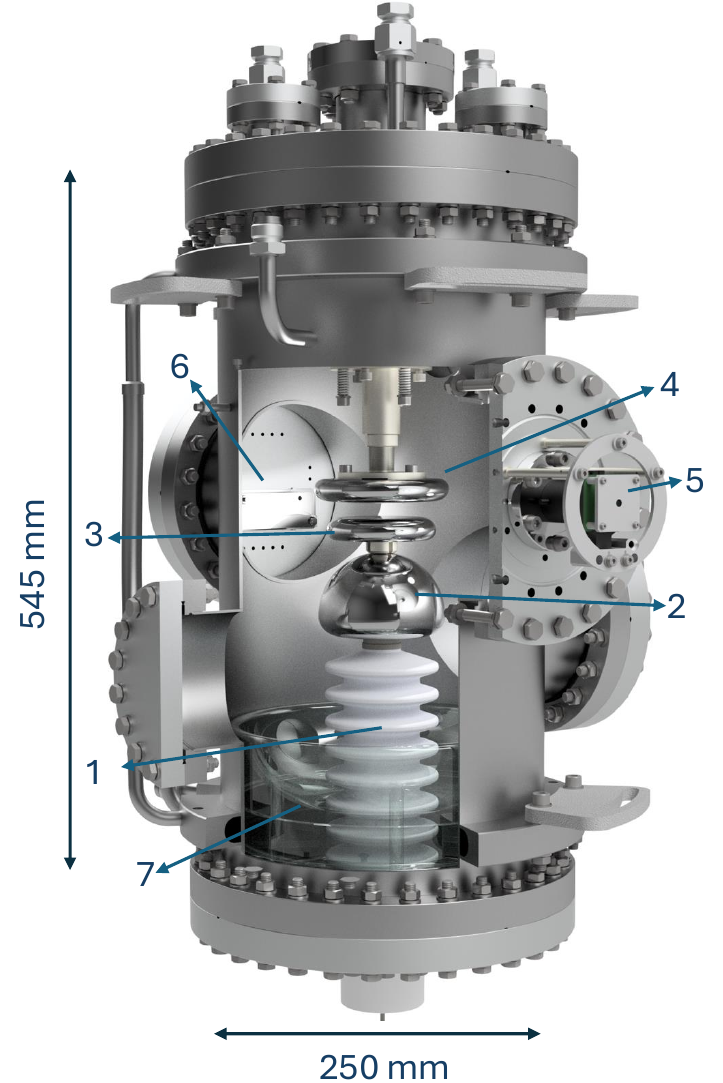}
    \end{subfigure}
    \begin{subfigure}[b]{0.44\textwidth}
         \centering
         \includegraphics[width=\textwidth]{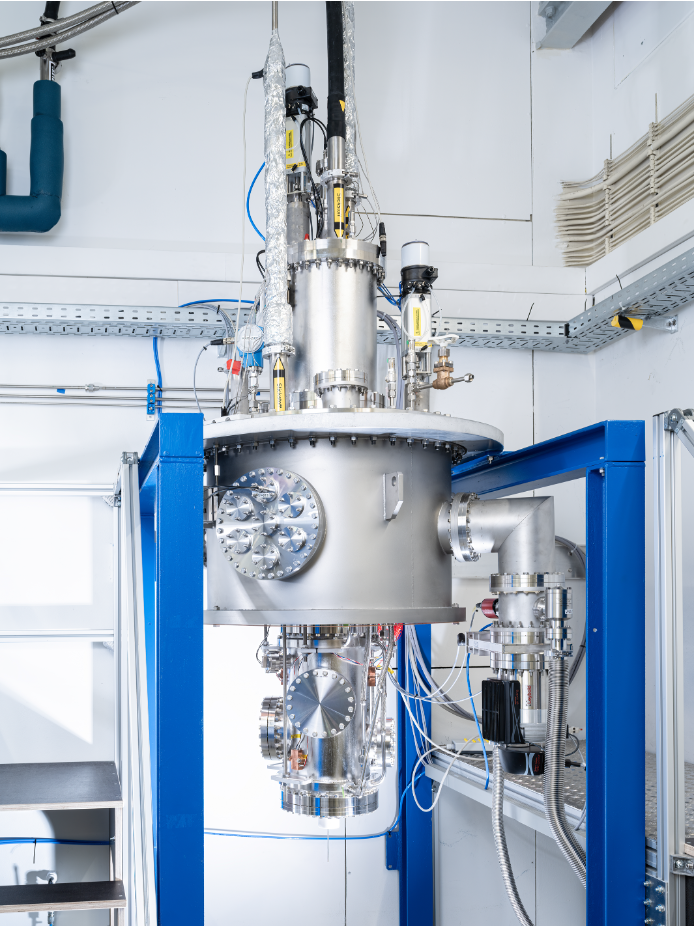}
    \end{subfigure}
\caption{Left: Construction drawing of the MOTION cryostat inner vessel. Inside the inner vessel, an HV feedthrough (1) is terminated (2) and connected to the cathode electrode (3). Another electrode (4), at ground potential, faces it. The space between the two Rogowski-profile electrodes is observed with a camera (5), through a DN40CF viewport. A liquid level readout (6) indicates whether both electrodes are submerged in LXe. A filler (7) displaces LXe towards the volume around the electrodes. Right: Picture of the MOTION cryostat. Credit: A. Bramsiepe / KIT.}
\label{fig:motion}
\end{figure}

Inside the inner vessel, two electrodes are immersed in LXe, where electric fields around \SI{-200}{kV/cm} can be tested. Figure~\ref{fig:motion} shows a render of the MOTION cryostat. The electrode-only configuration consists of two solid electrodes facing each other: a negatively biased cathode and an anode at ground potential, separated initially by a \SI{10}{mm} gap, with diameters of \SI{60}{mm} and \SI{80}{mm}, respectively. The profile of these electrodes was optimised based on~\cite{trinh_electrode_1980} to obtain a uniform electric field between them. The cathode voltage can be ramped up to \SI{200}{kV}, where the capacitive (transient) and resistive (stationary) regimes can be studied through the monitoring of the current in the anode and the current and voltage in the FuG Elektronik HCP 140 power supply, which provides HV of up to \SI{200}{kV} DC (negative polarity) with a maximum current of \SI{0.7}{\milli\ampere}. A mechanical joint can control the anode position and alignment, and a 2-axis inclinometer monitors this alignment (Murata, SCL3400-D01). In an initial stage, the HV feedthrough utilised is a commercial one (Hositrad, 6722-01-CF). Additionally, a camera on the vacuum side of the cryostat, facing a viewport on that side, can detect any prior electroluminescence that accompanies a dielectric discharge~\cite{franzke_micro-discharge_2009} in LXe, or bubbling due to heat dissipation. A second configuration, in which the electrodes are replaced by a dual-phase TPC, is under design and will follow the electrode-only phase.

\subsection{Level probe}

\begin{figure}[!h]\centering
   \centering
    \begin{subfigure}{0.39\textwidth}
        \includegraphics[height=5.5cm]{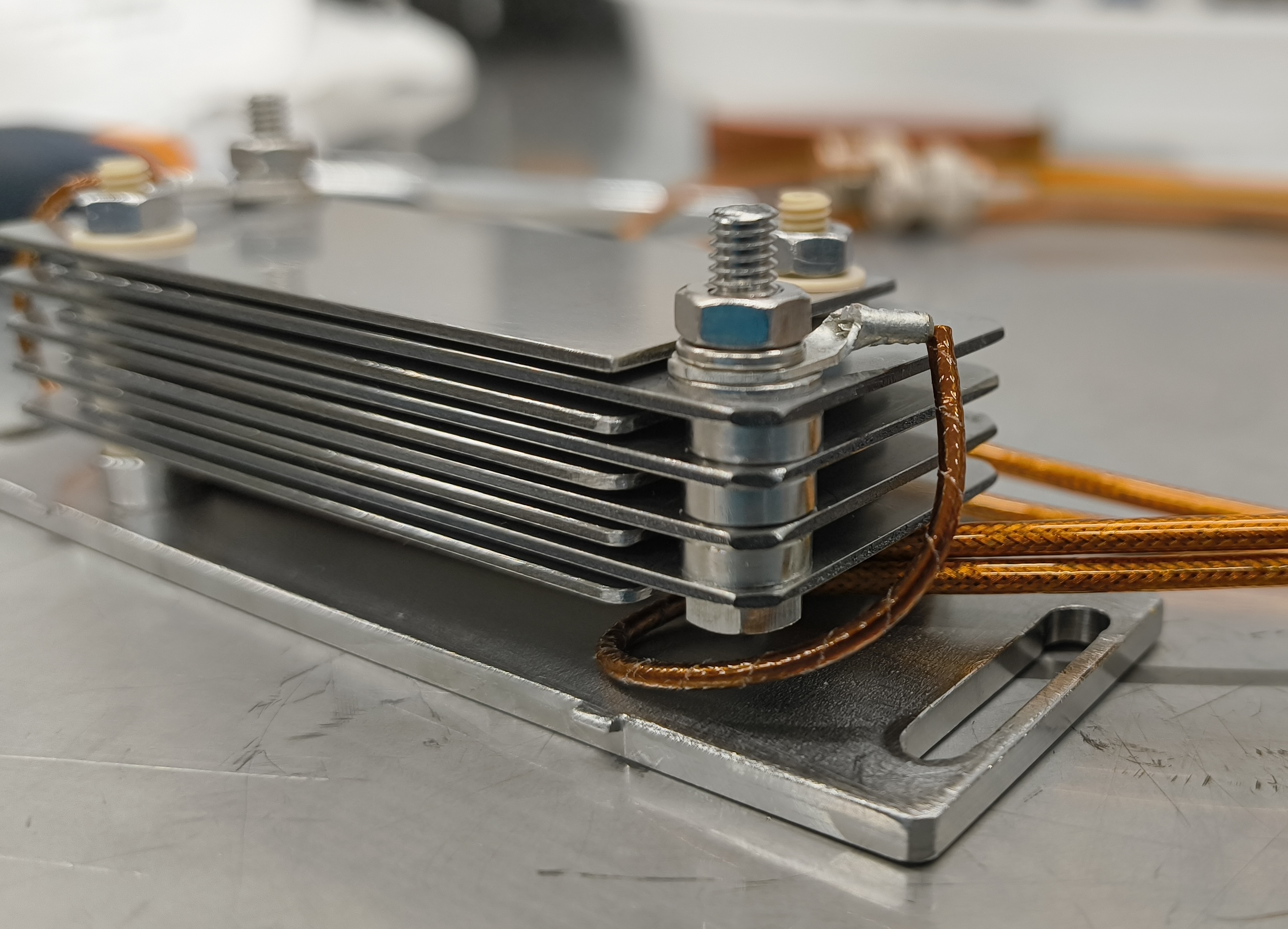}
    \end{subfigure}
    \hfill
    \begin{subfigure}{0.6\textwidth}
        \includegraphics[height=5.5cm]{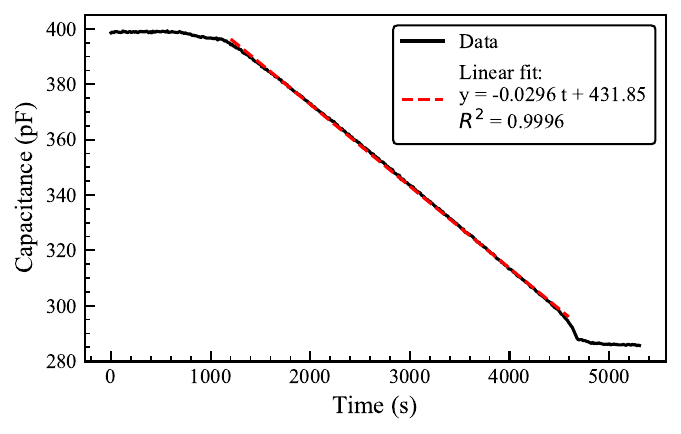}
    \end{subfigure}
    \caption{Left: stacked-plate capacitive xenon level probe with adjustable bulk capacitance. Right: level probe capacitance readout as a function of time; the linear fit indicates a constant evaporation rate.}
  \label{f:level}
\end{figure}

The xenon level probe consists of a stack of stainless steel plates that form a capacitive sensor, as shown in the left panel of Figure~\ref{f:level}. The probe's bulk capacitance can be adjusted by varying the number of plates in the stack. Adjacent plates are alternately connected to the two electrodes of the level transducer using two coaxial cables. The plate assembly is enclosed in a grounded stainless steel shield to minimize external electromagnetic interference. Each overlapping plate section measures \SI{60}{\milli\metre} $\times$ \SI{30}{\milli\metre}. Glass-filled PEEK washers minimize thermal contraction during cooldown, preserving the plate spacing and ensuring a stable capacitance response.

The probe was characterized by immersing it in LN$_2$. A constant-power heater was used to evaporate the LN$_2$, causing the liquid level to decrease at a constant rate, as shown on the right in Figure~\ref{f:level}. A linear response is observed over most of the probe's active range. Deviations from linearity near the ends of the probe are attributed to capillary effects, which distort the liquid surface around the sensing plates. This measurement demonstrates the proper operation of both the level probe and the capacitance transducer. An in-situ calibration is nonetheless required after installation, since small mechanical shifts in probe geometry and parasitic capacitance affect the absolute capacitance and hence the level calibration.

The first version of the capacitance transducer was based on the Smartec UTI capacitance-to-digital converter~\cite{UTI}, which provides a measurement range of up to \SI{300}{\pico\farad}. The circuit uses a precision \SI{100}{\pico\farad} reference capacitor and provides a resolution of 14~bits. The UTI is mounted on an evaluation board equipped with a microcontroller and an I$^2$C interface. An Arduino microcontroller reads the measurement data via I$^2$C and forwards it to the slow-control system via RS485 protocol. The UTI employs a charge-amplifier measurement principle, which effectively eliminates the influence of the cable capacitance on the measured probe capacitance~\cite{UTI}.

The second-generation transducer is based on the Analog Devices AD7746 capacitance-to-digital converter~\cite{cdc}, which integrates the functionality of the previous design into a single integrated circuit. The device employs a $\Sigma$--$\Delta$ modulator and, like the UTI, compensates for the capacitance of the connecting cables. The AD7746 provides a programmable capacitance offset that can be used to compensate for the probe's bulk capacitance. Its nominal input range is \(\pm\SI{4}{\pico\farad}\) (\SI{8}{\pico\farad} differential), which can be extended by adding an external operational amplifier. The converter performs continuous self-calibration using an internal reference and achieves a resolution of 21~bits. The original level probe was designed with a relatively small plate spacing to maximize the capacitance for the first-generation transducer. The AD7746's substantially higher resolution allows the plate spacing to be increased while maintaining sufficient sensitivity; the larger spacing in turn reduces capillary effects, improving linearity and accuracy. 

\subsection{Pre-breakdown current readout}
A transimpedance amplifier (TIA), which converts small input currents into a measurable voltage, is used to measure pre-breakdown currents in the electrode gap. A FEMTO DLPCA-200~\cite{TIA} variable-gain, low-noise current amplifier was characterized in low-noise mode. A known waveform was applied to the TIA through a resistor-loaded BNC cable, and the TIA output was measured using an oscilloscope. The gain linearity was verified over the full input current ranges specified in the datasheet, from \(\pm\SI{10}{\milli\ampere}\), to \(\pm\SI{0.01}{\micro\ampere}\) for gain settings from \(\SI{e3}{\volt\per\ampere}\) to \(\SI{e9}{\volt\per\ampere}\). The bandwidth was investigated by varying the input signal frequency and measuring the output voltage. At the highest gain setting of \SI{e9}{\volt\per\ampere}, the response remained linear up to \SI{1}{\kilo\hertz}. At lower gain settings, linearity was maintained over higher frequencies. 

The absolute maximum rating of the DLPCA-200 permits discharges of up to \SI{3}{\kilo\volt} from a \SI{200}{\pico\farad} source. Since the planned electrode test setup significantly exceeds these limits in both capacitance and voltage, a Transient Voltage Suppressor (TVS) protection board will be implemented to prevent both catastrophic failure and subtler degradation effects such as nonlinear behaviour. During a discharge event, the TVS diode clamps the voltage and diverts current to ground, thereby protecting the amplifier.

\subsection{Camera system}

Visual diagnostics are essential for understanding HV behavior in LXe by providing real-time observations of the electrode region that complement electrical measurements. The imaging system consists of a Raspberry Pi Camera Module 3 featuring a 12 MP Sony IMX708 CMOS sensor with a native resolution of $4608 \times 2592$ and a \SI{1.4}{\micro\meter} pixel pitch, coupled to a \SI{16}{\milli\meter} C-mount telephoto lens to provide high spatial resolution of the electrode gap. The camera enables direct correlation of electrical signals with physical events, confirming genuine breakdowns and monitoring the electrode gap during high-voltage ramps. It also verifies bubble-free liquid conditions before testing, captures precursors such as micro-arcs, records post-breakdown bubble dynamics, and detects optical disturbances caused by localized heating in the LXe. 

The outgassing of the Raspberry Pi camera setup was tested over a pressure range of approximately \SIrange{1e-5}{1e-4}{\milli\bar}. The measured outgassing rate was consistent with the expected values for electronic components operated under vacuum conditions. The flange has the option to mount an additional tube that isolates the volume surrounding the camera from the rest of the vacuum system, in case the outgassing is not well tolerated by the system. This is mounted over a plate with good thermal conductance, while heaters and temperature sensors monitor that the temperature does not drop in ranges where the operation of the camera is known to degrade~\cite{DUNE:2026mze}.

\section{Slow control system}

The slow control system provides continuous monitoring and control of the detector during commissioning and physics operation. It supervises detector parameters, including temperatures, pressures, liquid level, flow rate, valve states, and heater operation. Measurements acquired from the sensors are processed in LabVIEW~\cite{labview} for real-time monitoring, alarm handling, and valve control. The data are subsequently archived in an InfluxDB~\cite{influxdb} time-series database and visualized using Grafana~\cite{grafana} for both real-time monitoring and long-term analysis.

\begin{itemize}
\item \textbf{System hardware:} The instrumentation is interfaced through two National Instruments (NI) CompactRIO (cRIO)~\cite{crio} systems connected via an Ethernet network. The cRIO controllers acquire analogue signals from sensors and provide control outputs for process control components such as pneumatic valves. The installed NI I/O modules provide interfaces for resistance temperature sensors (PT100 and PT1000), analog current signals (\SI{4}{\milli\ampere}--\SI{20}{\milli\ampere}), analog voltage signals (\SI{0}{\volt}--\SI{10}{\volt}), and digital input/output channels used for detector monitoring and valve control. The modular cRIO architecture allows straightforward expansion with additional I/O modules to support future detector upgrades.

\item \textbf{LabVIEW monitoring software:} The monitoring software is developed using LabVIEW and runs on a Windows workstation that serves as the main operator interface. LabVIEW communicates with the cRIO controllers over an Ethernet network, continuously acquires all sensor data with a sampling period of \SI{1}{\second}, and logs the data locally on the workstation. The workstation is assigned a fixed IP address and is accessible only within the local network. The acquired data are displayed in real time and simultaneously forwarded to the InfluxDB database for permanent storage. The graphical user interface also allows operators to manually open and close control valves and continuously evaluates user-defined alarm conditions. Whenever a monitored parameter exceeds its predefined threshold, an alarm is generated to notify the operator.

\item \textbf{Data storage} InfluxDB~2.0 is used as the central repository for all slow control data. Measurements acquired by the cRIO controllers and processed by LabVIEW are written to the database together with precise timestamps. This enables efficient storage and retrieval of detector operating parameters over arbitrary time intervals and provides a complete history of detector operation for offline analysis and diagnostics.

\item \textbf{Data visualization} Grafana is used exclusively for visualization of the data stored in InfluxDB. It accesses the database directly to generate interactive dashboards for both real-time and historical monitoring, with database connections restricted to authorized clients only. Operators can inspect trends, compare multiple sensor parameters, and review the detector operating history through customizable plots. An example of the Grafana monitoring interface is shown in Figure~\ref{f:grafana}.

\begin{figure}[!t]\centering
  \includegraphics[width=\linewidth]{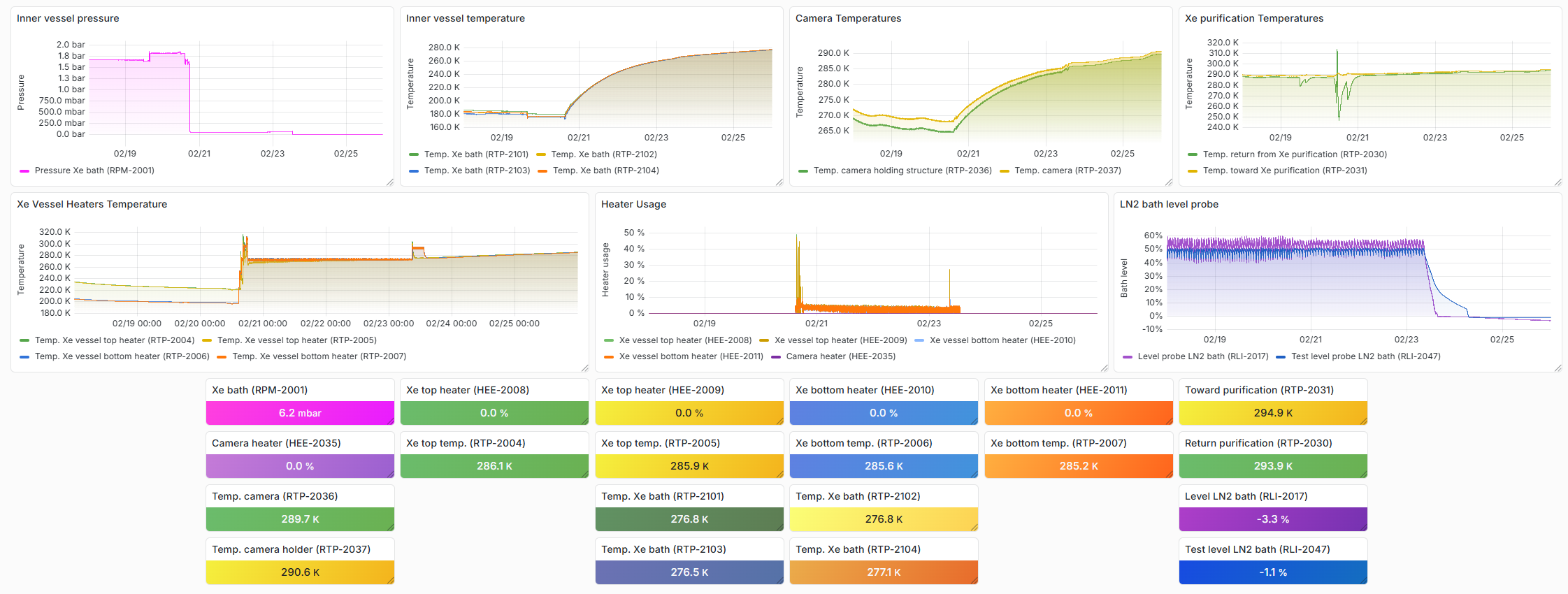}
  \caption{Overview of the Grafana-based slow-control dashboard developed for monitoring the inner vessel.}
  \label{f:grafana}
\end{figure}

\end{itemize}

\section{Gas system commissioning}

The first operation of MOTION with LXe took place in early 2026 and included a dedicated commissioning phase, with a partial fill of xenon inside the system. The phase aimed to demonstrate stable system operation under LXe conditions and included xenon liquefaction, recirculation at different flow rates through the filters and getter, and recuperation. The resulting measurements were used to determine the maximum recirculation flow rate achievable with the current MOTION setup, which affects the maximum liquid xenon purity that can be reached and maintained. 

\subsection{Xenon transfer}

Before xenon was transferred to the system, the inner vessel was evacuated. Throughout the commissioning period, the mean pressure in the outer vessel was \SI{1.1 \pm 0.2e-5}{\milli\bar}, which was concluded to be sufficiently low to minimize thermal conductance through air. Leak testing of the full system exhibited an integral leak rate of \SI{2e-7}{\milli\bar\liter\per\second}. 

The system was filled with xenon over three filling periods. During the first period, approximately \SI{600}{\gram} of GXe was injected to test the filling procedure. However, the liquefaction and subsequent evaporation of the xenon exceeded expectations, causing a rapid pressure increase in the system. To maintain safe operating conditions, approximately \SI{400}{\gram} of xenon was transferred from the inner vessel to the emergency vessels. This demonstrated that the emergency lines and vessels meet design expectations. Across the second and third filling period, a total of approximately \SI{400}{\gram} of GXe was injected, with approximately \SI{200}{\gram} added during each period, ultimately bringing the inner vessel pressure to \SI{1.8}{\bar}.

\subsection{Recirculation}

\begin{figure}[!htbp]
    \centering
    \includegraphics[width=0.49\linewidth]{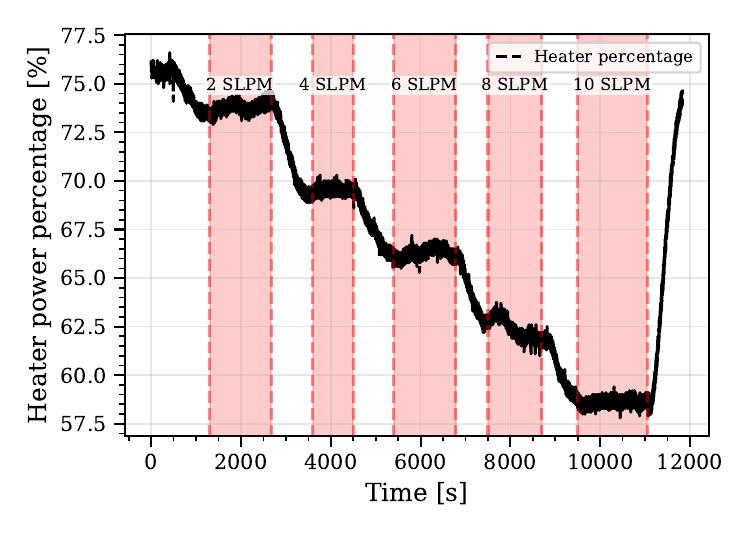}
    \includegraphics[width=0.49\linewidth]{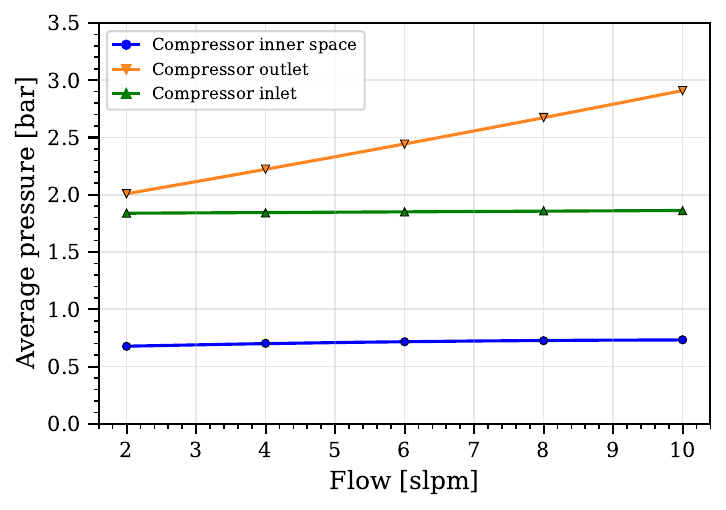}
    \caption{Left: The cold head heater power percentage as a function of time, zoomed to the \SIrange{2}{10}{\SLPM} range. The shaded red regions indicate the selected time intervals for each flow rate, chosen where the heater power is approximately stable.
    Right: The average pressure measured by each sensor as a function of flow rate, calculated over the selected time intervals shown in the figure on the left. The blue circle, orange downward triangle, and green upward triangle data points correspond to the compressor inner space, compressor outlet, and compressor inlet, respectively.}
    \label{fig:heater_power/recirculation_ressure}
\end{figure}

After the filling of LXe into the inner vessel, recirculation through the purification panel was tested with the getter in operation, as the getter was expected to provide a large flow impedance in the recirculation loop due to the filter inside. The flow rate, measured in standard liter per minute (SLPM), was increased in steps of \SI{2}{\SLPM}, starting from \SI{0}{\SLPM}, and was controlled using the mass flow controller. After each adjustment, the heater on the cold head was monitored until stable conditions were observed before proceeding to the next flow rate, as shown on the left in Figure~\ref{fig:heater_power/recirculation_ressure}. Heating to the condenser is provided by two Lake Shore HTR-50 cartridge heaters, wired in parallel and driven at \SI{48}{\volt} each. The heaters are inserted into radial holes on opposite sides of the copper cold head flange to enhance uniform heating. 

The red-shaded regions in the left panel of Figure~\ref{fig:heater_power/recirculation_ressure} indicate the selected time intervals for each flow rate under approximately stable heater conditions. As the GXe recirculated at each flow rate, the gas pressure was measured at several points relative to the compressor. The pressure of the gas coming from the cryostat was measured by a pressure sensor situated just before the flow meter and is referred to as the `compressor inlet' pressure. Using the same sensor type, the pressure in the `compressor inner space' is measured. The pressure of the gas leaving the compressor is again measured by the same sensor type, and is referred to as the `compressor outlet' pressure. Using the time intervals specified in the left panel of Figure~\ref{fig:heater_power/recirculation_ressure}, the average pressure for each sensor is calculated within each time interval. This results in the average pressure measured by each sensor as a function of flow rate, as shown in the right panel of Figure~\ref{fig:heater_power/recirculation_ressure}.

At a flow rate of \SI{10}{\SLPM}, the gas pressure at the compressor outlet reaches \SI{2.9}{\bar}. The compressor is rated up to \SI{3}{\bar}~\cite{KNF}; therefore, \SI{10}{\SLPM} is the maximum recirculation flow rate achievable with the current setup. 

\subsection{Temperature stability}

\begin{figure}[!htbp] 
    \centering    
    \includegraphics[width=0.6\textwidth]{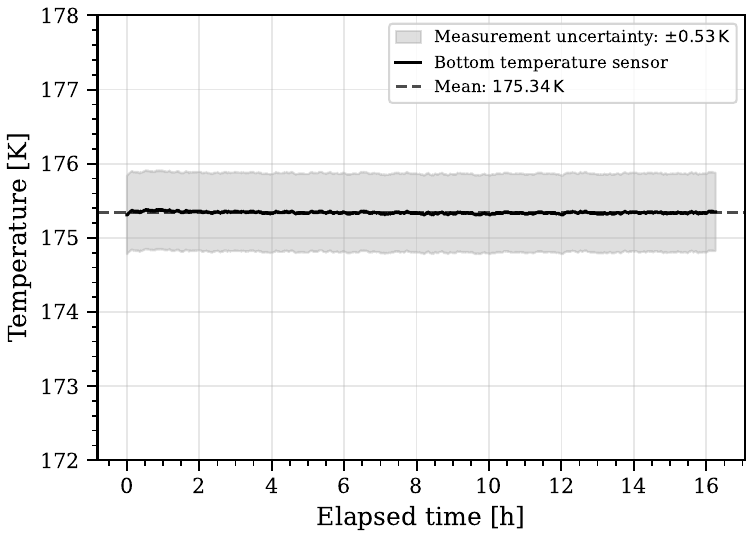}
    \caption{Temperature measured by the bottommost temperature sensor inside the inner vessel as a function of the elapsed monitoring time. The elapsed time is selected to start one hour after the third filling period, allowing the inner vessel temperature to adjust to the setpoint temperature.}
    \label{fig:temperature_stability}
\end{figure}

The temperature stability of the inner vessel was monitored after the third xenon filling period, using the bottommost temperature sensor installed inside the vessel. Following the filling, the cold-head setpoint temperature was raised from \SI{174}{\kelvin} to \SI{176}{\kelvin}, and the inner vessel pressure was measured to be \SI{1.83 \pm 0.03}{\bar}. The stability-monitoring period was defined to begin one hour after the end of the third filling period, allowing time for the system to adjust to the setpoint temperature. The temperature was then monitored for approximately 16 hours,  after which xenon recuperation began. The measured temperature as a function of the elapsed monitoring time is shown in Figure~\ref{fig:temperature_stability}.

The mean temperature over the selected period was \SI{175.34 \pm 0.53}{\kelvin}, where the uncertainty is the combined systematic measurement uncertainty of the temperature sensor and the associated readout module used. As seen in Figure~\ref{fig:temperature_stability}, the measured temperature remained stable throughout the full 16-hour monitoring period.

\subsection{Recuperation}

The xenon recuperation procedure was also tested. By cryogenically cooling the storage bottles with liquid nitrogen, approximately \SI{600}{\gram} of GXe was recovered from the inner vessel in under 30 minutes. Recovery continued until the pressure reached approximately \SI{10}{\milli\bar}, demonstrating that the xenon can be effectively recuperated.


\section{Electrode surface study}

Local field enhancement at asperities is suspected to drive breakdown. To develop a protocol for studying the surfaces of HV components, simple stainless steel disks with a diameter of \SI{40}{mm} were machined in our local mechanical workshop. The surface finish of the electrodes was characterized on a Keyence VK-X3050 microscope~\cite{keyence}. The laser microscope, which uses the focal principle, is used for our surface roughness measurements. Laser microscopes provide quick, non-contact surface roughness measurements as numerical data and can measure small grooves without damaging the sample surface. Because the confocal microscope maximizes reflected-light intensity only when the target is in focus, its performance depends heavily on accurately detecting this peak. A small pinhole (several tens of \si{\micro\metre}) in front of the detector suppresses out-of-focus light, enabling precise height measurements. The profiler acquires \num{1024} (X) ($\times$) \num{768} (Y) pixels. The reflected light intensity and the lens height position are recorded in memory for each pixel. When scanning one surface finishes, the objective lens moves 100 times in the Z direction for every pixel. The microscope achieves a Z-axis measurement resolution of \SI{1}{\nano\metre}.

A stylus measurement was first performed using a perthometer (Mahr M2) to examine the mechanical reference roughness of the machined electrode surface (reference value: \SI{0.2}{\micro\metre}). The measured surface roughness is \SI{0.373}{\micro\metre} with a travelling distance of \SI{5.6}{\milli\metre}. The same location~\cite{pradhan_surface-topography_2025} was subsequently measured using the stylus mode of the Keyence microscope at $20\times$ magnification. Multi-line surface roughness measurements were performed according to ISO 21920, yielding a comparable surface roughness value of \SI{0.349}{\micro\metre}. 

\begin{figure}[!htbp]\centering
  \includegraphics[width=0.8\linewidth]{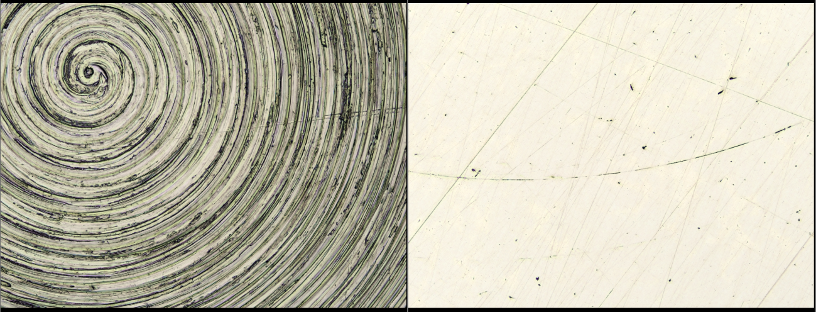}
  \includegraphics[width=0.8\linewidth]{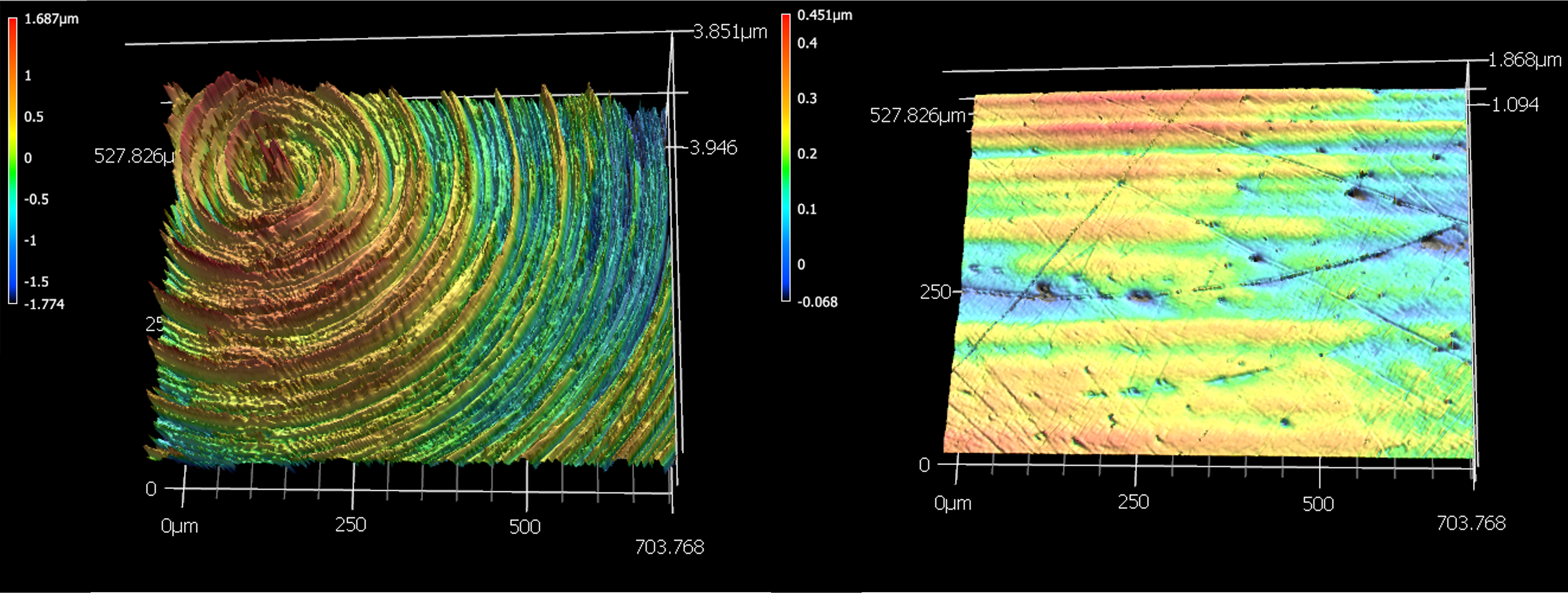}
  \caption{Surface morphology of the same area on the same electrode before (left) and after (right) polishing. The top row shows combined laser and optical microscope images, while the bottom row presents the corresponding 3D surface topography with a vertical height magnification of \SI{5000}{\percent}.}
  \label{f:PnL}
\end{figure}

A $20\times$ objective magnification was later selected for the machined electrode surface study, as it provides sufficient resolution and an appropriate field of view under combined optical and laser imaging. The same scanning procedure was subsequently applied to the polished electrode to enable a direct comparison of the same surface area before and after polishing. Figure~\ref{f:PnL} compares the surface morphology of the same area on the same electrode before and after polishing. The left column shows the lathed surface, while the right column shows the polished surface. The lathed surface exhibits characteristic periodic machining marks produced by the turning process, whereas the polished surface displays randomly oriented polishing tracks. The top row presents the combined laser and optical microscope images, and the bottom row shows the corresponding 3D surface topography with a vertical height magnification of \SI{5000}{\percent}. The measured surface height ranges from \SI{-1.774}{\micro\metre} to \SI{1.770}{\micro\metre} for the lathed surface and from \SI{-0.068}{\micro\metre} to \SI{0.451}{\micro\metre} for the polished surface, indicating a substantially smoother surface after polishing.

The motorized XY stage, with a travel range of \SI{100}{\milli\metre} $\times$ \SI{100}{\milli\metre}, automatically moves the sample according to the predefined scan area. Since the electrode has a diameter of \SI{40}{\milli\metre}, the entire surface can be scanned automatically using the teaching mode, which supports up to 3000 registered locations for sequential image acquisition. At $20\times$ magnification, each image (tile) covers an area of approximately \SI{703}{\micro\metre} $\times$ \SI{527}{\micro\metre}. Scanning the entire electrode requires two consecutive measurements totalling \num{3372} registered locations; the complete scanning process takes approximately \SI{3}{\day}.

For the polished surface, a magnification of $50\times$ is required to resolve finer surface features. However, this higher magnification reduces the field of view of each tile to approximately \SI{277}{\micro\metre} $\times$ \SI{208}{\micro\metre} and increases the acquisition time per tile due to the higher imaging resolution. Consequently, a total of 21795 tiles are required to cover the entire surface, resulting in a total scanning time of more than \SI{15}{\day}. This long acquisition time presents a significant challenge, as the measurement cannot be easily interrupted and resumed; the electrode must remain in the exact same position throughout the entire scanning process to ensure accurate tile registration and consistent surface mapping.

The surface texture analysis was performed using the VK-X3000 Series Multifile Analyzer software. Since the software supports a maximum of 10 analysis lines or areas per tile, each tile was divided into a $5 \times 2$ grid of subregions to maximise spatial coverage and ensure complete pixel utilisation. An analysis template was first created from a single tile and then applied to all remaining tiles using the batch analysis function. In accordance with ISO~25178, each subregion was levelled, followed by the application of an S-filter with a nesting index equal to three times the planar resolution and an L-filter with a nesting index equal to one-fifth of the profile length, following the manufacturer's recommendations~\cite{keyence_filter}. After these preprocessing steps, the areal surface texture parameters, such as the maximum height ($S_z$), could be acquired for every subregion through batch analysis.

\begin{figure}[!t]\centering
  \includegraphics[width=0.49\linewidth]{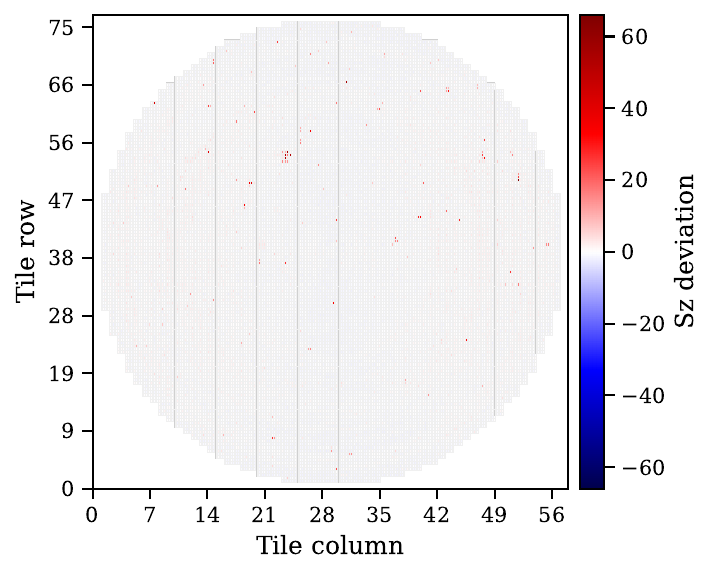}
  \includegraphics[width=0.49\linewidth]{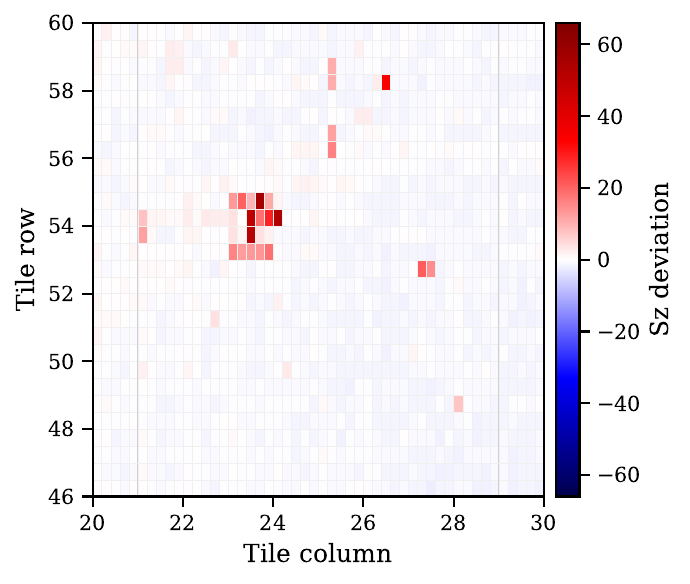}
  \includegraphics[width=0.6\linewidth]{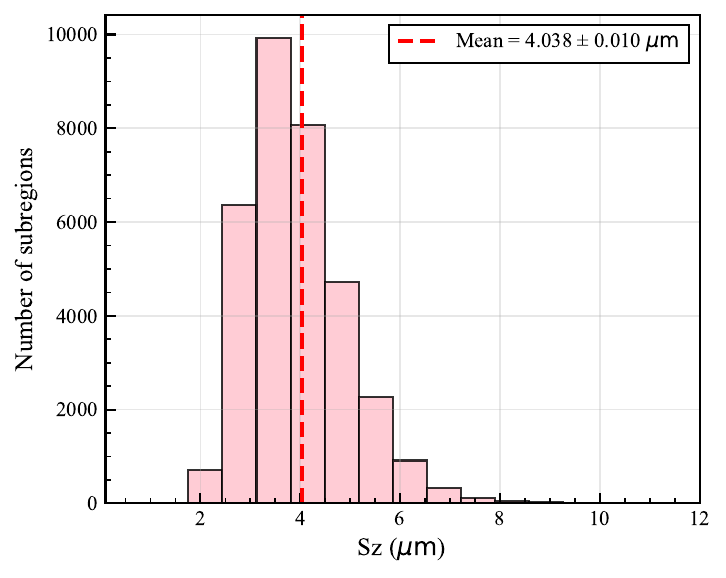}
  \caption{Left: Spatial map of the tile-averaged $S_z$ deviation from the mean $S_z$ of all tiles on the lathed surface. Right: Magnified view highlighting a potential surface defect. Bottom: Distribution of $S_z$ values measured for all subregions (10 subregions per tile).}
  \label{f:Sz}
\end{figure}

Figure~\ref{f:Sz} shows the deviation of $S_z$ from the average value (bottom) of all tiles for the lathed surface at $20\times$. While the lathed topography is dominated by periodic machining marks (left), the deviation map demonstrates a visualization approach that highlights localized variations after removing the tile-average value (right). This method will be applied to the polished surfaces, where subtle spatial variations are expected to be more informative. 

Full-surface scanning of the polished electrode is planned. Heat maps of the areal surface texture parameters, such as arithmetic mean height ($S_a$), $S_z$, skewness ($S_{sk}$), and kurtosis ($S_{ku}$)~\cite{3Dparameters}, will be generated to evaluate the spatial variation in surface topography. A systematic study of dielectric breakdown in a dual-phase LXe detector under various controlled conditions will be conducted, supported by finite element simulations. These simulations will focus on identifying the physical and geometrical factors that initiate and propagate electrical discharges, including electrode geometry, surface features, regions of local electric-field enhancement, and material residues on electrode surfaces.

\section{Ongoing upgrade and planned measurements}

\begin{figure}[!htbp] 
    \centering    
    \includegraphics[width=0.35\textwidth]{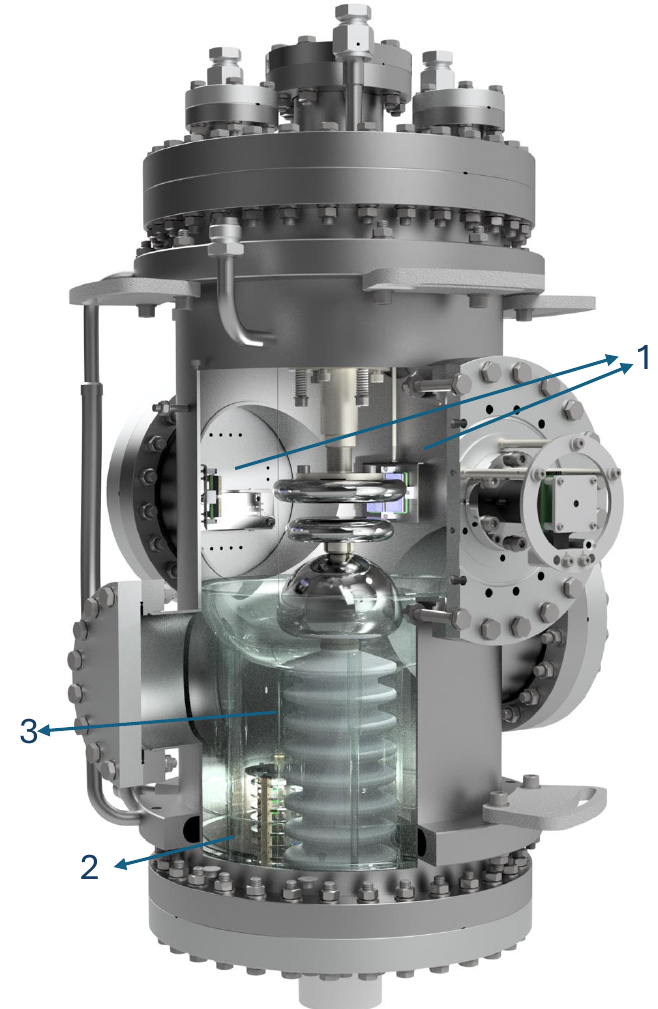}
    \caption{Upgraded vessel with (1) two SiPMs observing the region with higher electric field around the HV termination and electrodes, (2) a PM that provides information about the LXe purity, and (3) a thin tube providing a liquid level sensor based on differential pressure}
    \label{fig:improved_detector}
\end{figure}

After characterising the gas system and instrumenting the main components inside the detector, the detector is currently undergoing an upgrade to add systems that will address unknowns about systematic effects of discharges in HV components and serve as a platform to test a custom-made HV feedthrough for the XLZD experiment. Figure~\ref{fig:improved_detector} shows the MOTION detector with the addition of two SiPMs observing the HV components, a purity monitor (PM) on the bottom of the detector, and the differential pressure system to measure the liquid level inside the detector, also known as a long level meter. These components are described in detail in the following subsections.

\subsection{SiPM characterization}

SiPMs are a strong alternative to photomultiplier tubes (PMTs) as photosensors for next-generation xenon-based dark matter experiments~\cite{wu_electrode_2026, baudis_SiPM}. They can also be used to monitor light signals associated with breakdown precursors and to improve the understanding of breakdown mechanisms in dual-phase liquid xenon detectors. 

\begin{figure}[!htbp]\centering
  \includegraphics[height=5.3cm]{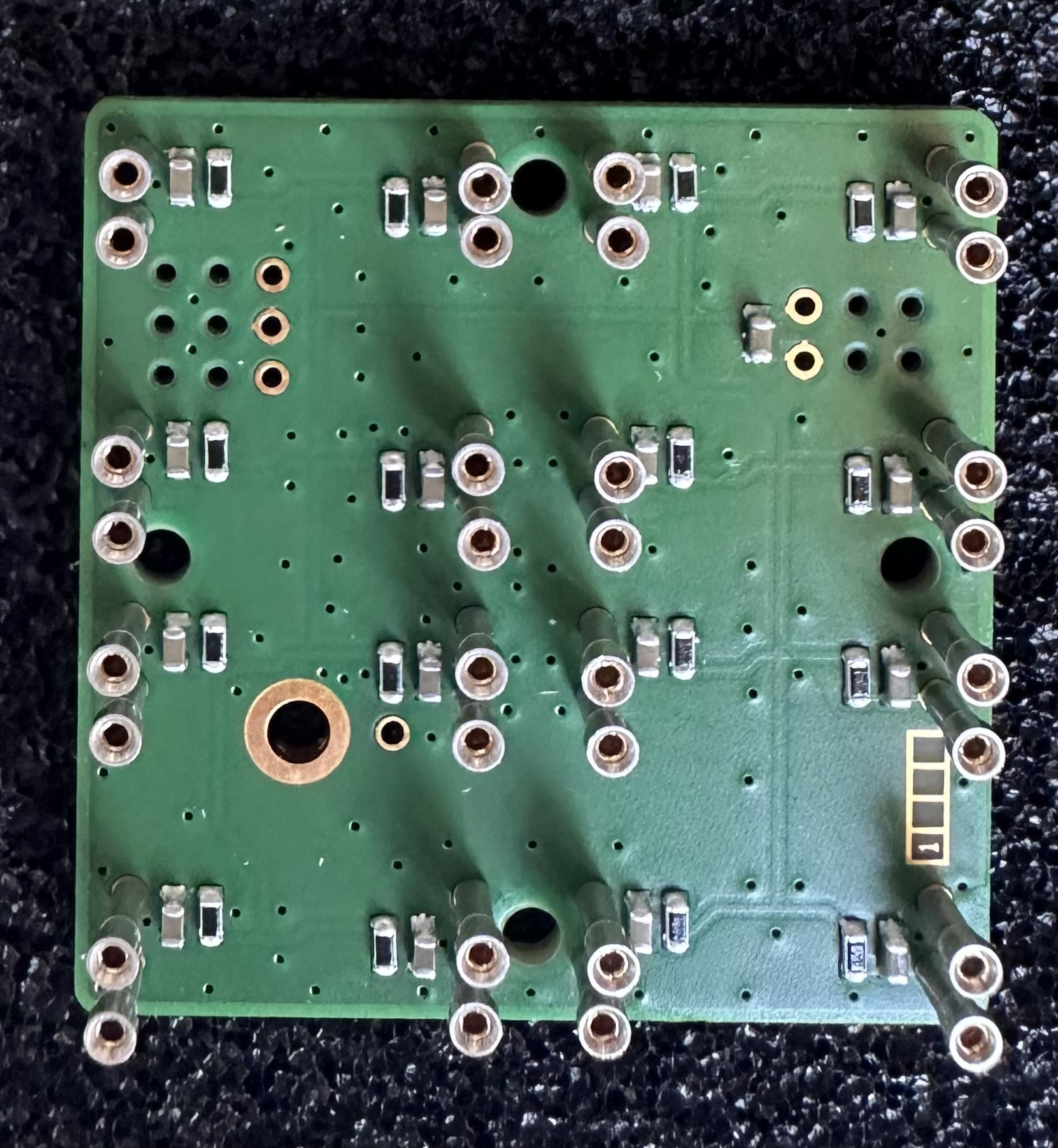}
  \includegraphics[height=5.3cm]{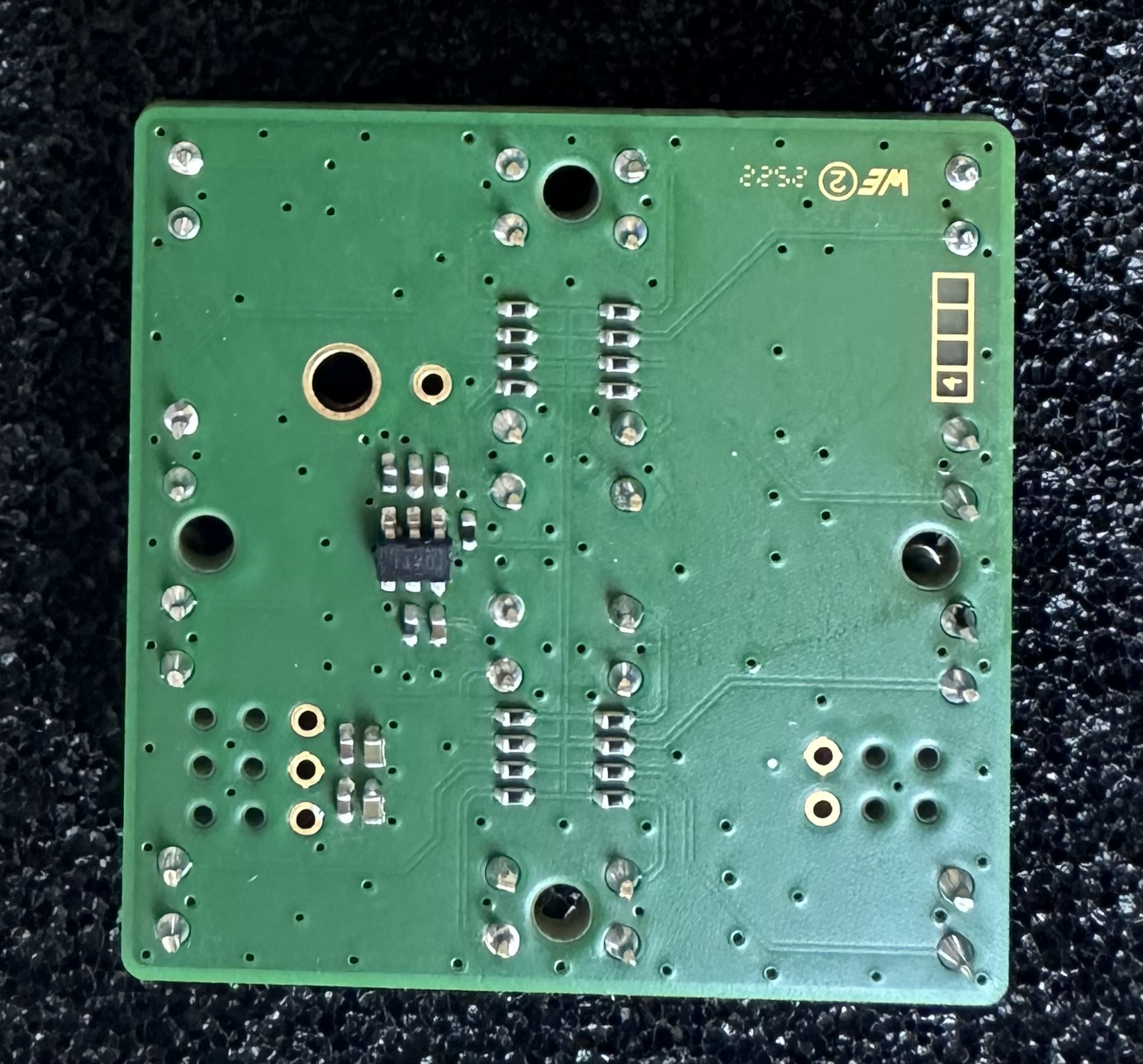}
  \includegraphics[height=5.3cm]{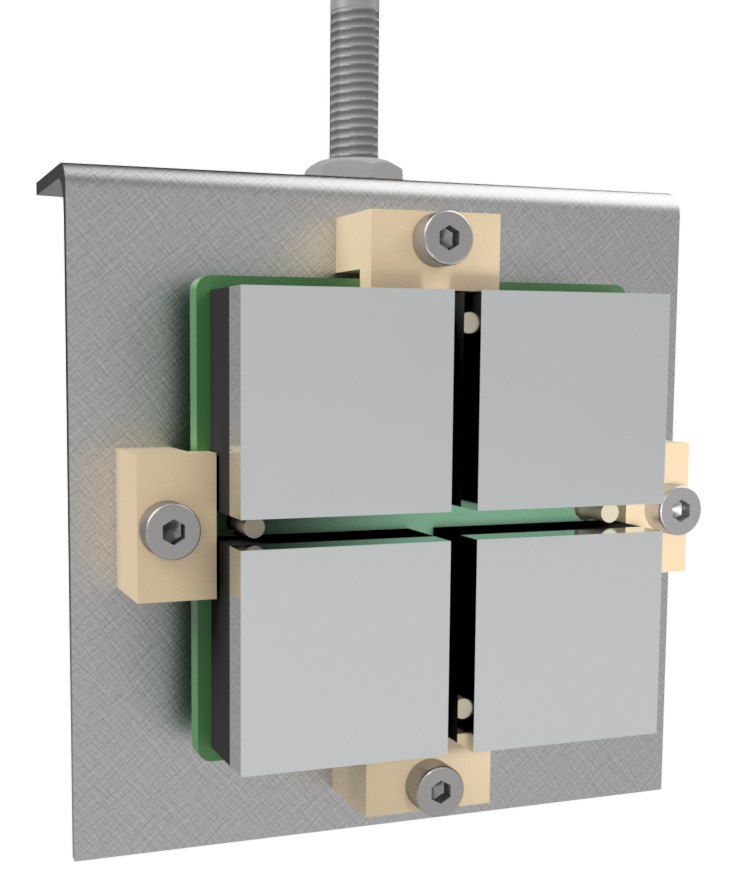}
  \caption {Overview of the \qtyproduct{34 x 34}{\milli\metre} four-layer SiPM board. The top and bottom views of the fabricated board are shown on the left and center, respectively, while the CAD model on the right illustrates the integration of the SiPM assembly within the inner vessel.}
  \label{f:sipm}
\end{figure}

In this work, we used a Hamamatsu S13371-6050CQ-02 SiPM, whose design is similar to that described in~\cite{SiPMVUV}. The developed SiPM board is shown in Figure~\ref{f:sipm}. The top and bottom views present the two sides of the fabricated printed circuit board. The accompanying CAD model depicts the mechanical implementation of the SiPM assembly inside the inner vessel. This VUV-sensitive SiPM has a photon detection efficiency of about $25$\% at \SI{178}{\nano\metre}, making it well suited for detecting LXe scintillation light.

While SiPMs offer advantages such as high radiopurity and fast pulse timing, they remain significantly inferior to PMTs in terms of dark count rate (DCR), even at low temperatures. At room temperature, the Hamamatsu SiPM exhibits a dark count rate on the order of a few Mcps. However, when cooled to \SI{180}{\kelvin}, the DCR decreases dramatically to about $\SI{1.82}{\hertz\per\square\milli\metre}$, resulting in an estimated total DCR of approximately \SI{65}{Hz} for the \qtyproduct{6 x 6}{\milli\metre} active area~\cite{Peres2023}.

The SiPM is being characterized in a custom-built chamber mounted on copper blocks cooled by liquid nitrogen and gaseous nitrogen flow. The temperature was monitored and regulated using a custom-built heater and PID controller. The single-ended output signal was amplified with a CAEN N979 fast amplifier~\cite{linear_amplifier} with a fixed gain of \qty{10} before being digitized by a 14-bit CAEN V1724~\cite{v1724} \SI{100}{\mega\text{S}\per\second} waveform digitizer. We aim to study the SiPM performance over a temperature range from \SI{160}{\kelvin} to \SI{210}{\kelvin}. The characterization includes measurements of gain as a function of bias voltage, breakdown voltage as a function of temperature, DCR as a function of gain and temperature, heat maps of DCR versus temperature and gain, single photoelectron (SPE) resolution as a function of gain, and crosstalk probability (CTP) as a function of gain. This characterization will provide the information necessary to evaluate the SiPM performance at the MOTION detector.

\subsection{Purity monitor}

Studying xenon purity is an important aspect of the systematic investigation of HV breakdown in liquid xenon. Commercially available xenon contains trace concentrations, typically on the order of $\sim$\si{\ppm}, of several impurities, which can degrade the signal by capturing UV photons and free electrons. In particular, the electronegative impurities such as  $\mathrm{O_2}$, $\mathrm{CO_2}$, and $\mathrm{N_2O}$, which get introduced through outgassing or small leaks, diffuse throughout the LXe volume, and subsequently capture free electrons~\cite{Hasterok_impurities}. Among these, oxygen is especially relevant because of its abundance in air and its high electron-attachment rate~\cite{attachment_rate}. The presence of electronegative impurities reduces the drift charge, degrading the measured charge signal, but may also decrease the probability of high-voltage breakdown~\cite{Xebra}. Therefore, constant work is done to reduce the presence of electronegative impurities. For this, gaseous xenon is circulated through hot zirconium getters~\cite{Dobi_2010}. However, some impurities may remain or get reintroduced into the system. Therefore, estimating the impurity concentration is crucial for detector performance and breakdown studies.

\begin{figure}[h!]
    \centering
    \begin{subfigure}[b]{0.30\textwidth}
         \centering
         \includegraphics[width=\textwidth]{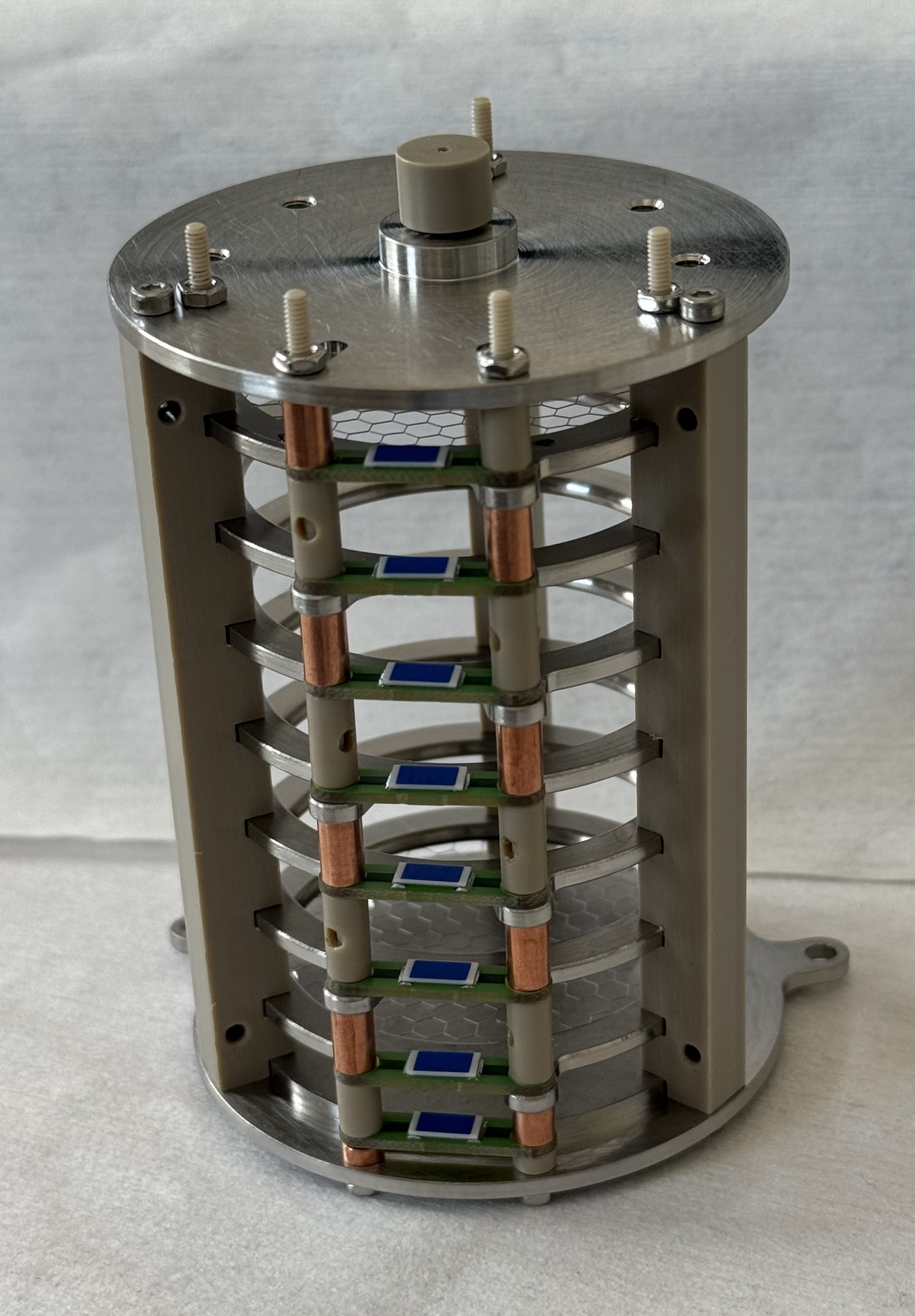}
    \end{subfigure}
    \begin{subfigure}[b]{0.40\textwidth}
         \centering
         \includegraphics[width=\textwidth]{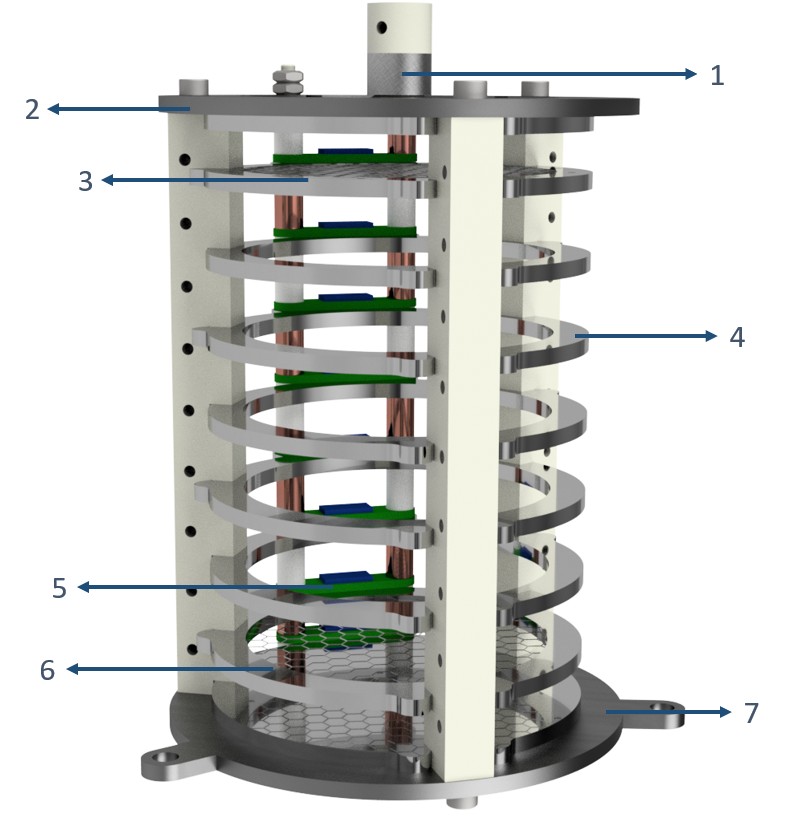}
    \end{subfigure}
\caption{The MOTION purity monitor. Left: Photograph of the built purity monitor. Right: Schematic view of the purity monitor. Legend: (1) Photocathode holder; (2) cathode disk; (3) cathode grid; (4) field-shaping rings; (5) resistor chain; (6) anode grid; (7) anode disk.}
\label{fig:PM}
\end{figure}

The concentration of electronegative impurities can be estimated by measuring the electron lifetime. For this purpose, dedicated PMs have been built based on the working principle of the ICARUS design~\cite{ICARUS_PM} and are utilized across current-generation experiments such as XENONnT~\cite{XenonNT_PM} and ProtoDUNE-SP~\cite{ProtoDUNE_PM}. The PM built for MOTION is approximately \SI{10}{\centi \meter} in height and is shown in Figure~\ref{fig:PM}.

Light pulses from a xenon flash lamp are delivered through an optical fiber to a gold-coated quartz photocathode, mounted in its holder. When illuminated with UV light, the gold layer emits electrons via the photoelectric effect. An applied electric field guides the emitted electrons through an opening in the cathode disk and towards the cathode grid. As the electrons drift from the cathode disk towards the cathode grid, their motion induces the first current pulse in the cathode readout. After passing through the cathode grid, the electrons drift through the field-shaping region towards the anode. Once they pass through the anode grid, their motion towards the anode disk induces a second current pulse, which ends when the electrons are fully collected on the anode disk.

The current signals are processed using custom-designed readout boards, based on the ones developed at the University of Zurich~\cite{Baudis2021-pr}. The readout boards provide the required HV bias and also filter, AC-couple, and amplify the signals before they are sent to the data acquisition. The electron lifetime is obtained from the ratio of the integrated anode and cathode current pulses and the measured electron drift time.  


\subsection{Planned measurements}

The arrangement of Rogowski electrodes monitored by a camera, two SiPMs, and a TIA in the current readout of the anode will allow measurement of dielectric breakdown inside LXe. In this measurement, the effects of bulk LXe dielectric strength and surface charge density are related, and imaging the electrodes before and after the test will enable a quality-assessment plan for conductors in the future XLZD detector. In this manner, surface coatings and chemical and mechanical treatments on electrodes, such as polishing and chemical passivation, can be quantitatively assessed for future use. While in the future the scanning of big surfaces, such as the ones required for \SI{3}{\metre} electrodes, remains not feasible, the analysis conducted here will be used to inform optimal surface treatments and coatings on smaller parts, before being applied in the large-scale assembly of HV components. The light readout from the SiPMs and cameras will provide information on electron emission from electrically stressed conductor surfaces, one of the mechanisms discussed in Section 1. A more detailed study of this effect will require upgrading the detector to a dual-phase TPC, foreseen after the electrode-only phase.

The purity monitor will provide additional information on the environmental properties of the LXe utilised in the detector and test the hypothesis that the presence of electronegative impurities diffused in the LXe can act as a suppressor of discharges. 

Additionally, this instrumentation allows for the testing of HV feedthroughs and cathode connections. The chamber can easily accommodate an HV feedthrough of the size expected for the XLZD detector, where its performance can be tested in LXe. HV feedthrough development for XLZD is currently underway, with the first component already in production. The design follows the approach adopted by the LZ, EXO-200, and DarkSide experiments and the nEXO proposal ~\cite{Mount:2017qzi,Auger:2012gs,Luzzi:2026bca,nexoHV}: rather than employing two separate feedthroughs (one at the air-to-vacuum interface and a second at the vacuum-to-GXe interface), the gaseous xenon volume is extended through an umbilical conduit, such that the HV cable passes through a single compression fitting and delivers the bias voltage continuously, without termination.

\begin{figure}[!htbp] 
    \centering    
\includegraphics[width=\textwidth]{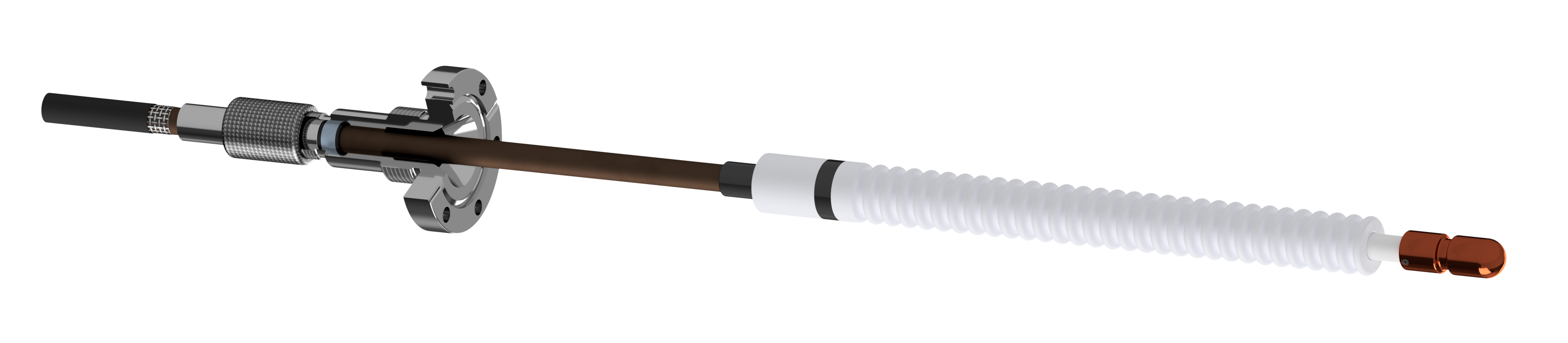}
    \caption{Schematic view of the HV cable passing through a compression fitting mounted on a CF40 vacuum flange, where the cable continues until the ground shielding is stripped away and in contact with a conductive polyethylene layer. The machined insulator, usually consisting of UHMWPE, increases the creepage distance and discharges to prevent surface flashover.}
    \label{fig:warm feedthrough}
\end{figure}

Figure~\ref{fig:warm feedthrough} shows the schematic view of the design of an HV feedthrough to be tested inside MOTION. The HV cable enters through a compression fitting, based on a set of two O-rings that provide a piston and radial seals, mounted on a DN40CF flange, continuing uninterrupted into the xenon volume. Downstream of the fitting, the outer ground shielding is stripped from the cable to expose the insulation, which is then brought into contact with a conductive polyethylene layer that grades the electric field at the shield termination. A machined insulator, typically made of ultra-high-molecular-weight polyethylene, surrounds this region, extending the creepage distance along the insulator surface and thereby suppressing surface flashover and spurious discharges. A threaded port on the flange allows an auxiliary vacuum volume to be attached and continuously monitored with a residual gas analyzer, enabling the detection of xenon leaks across the feedthrough seal.

\section{Summary and outlook }

MOTION provides a dedicated platform for systematic investigation of dielectric breakdown in liquid xenon. The facility integrates xenon storage, purification, cryogenics, and diagnostics to study breakdown mechanisms across a range of electrode geometries, surface conditions, and impurity levels.

The work presented here is part of a comprehensive R\&D program towards XLZD. Simultaneous electrical and optical measurements enable direct correlation of discharge precursors with breakdown events. Pre-breakdown current monitoring and high-resolution imaging of the electrode gap will characterize the dependence of breakdown field on surface finish, electrode geometry, and electronegative impurities. These expected results will help explain why current xenon TPCs fail to achieve their design voltages, including electrical breakdown well below the theoretical dielectric strength of the bulk liquid, as well as the emission of secondary electrons and luminescence from detector components.

Complementary diagnostics include SiPM-based optical monitoring of electroluminescence associated with discharge precursors, and a purity monitor to quantify electronegative impurity concentrations and test whether contaminants suppress or promote breakdown. Surface characterization via confocal laser microscopy will establish quality-assurance protocols for XLZD electrodes and conductors, with systematic assessment of surface treatments such as polishing and chemical passivation.

MOTION also serves as a testbed for XLZD HV feedthroughs. Full-scale prototypes of the compression-fitting design will be validated under realistic thermal and electrical stresses, with leak monitoring via residual gas analysis. These measurements clarify the electrostatic limits of xenon TPCs and identify mitigation strategies essential for next-generation dark matter detectors.

\acknowledgments

We thank Prof. Kathrin Valerius (KIT, Institute for Astroparticle Physics) for her continuous support, mentorship and valuable discussions. We are grateful to the engineering and technical staff for their excellent solutions and cooperative spirit, in particular Steffen Lichter, Michaela Meloni and Kshitija Satao (KIT, Institute for Astroparticle Physics). Among the scientific staff, we thank Thomas Höhn (KIT, Institute for Astroparticle Physics) for developing the cRIO and LabVIEW monitoring system, Sascha Wüstling (KIT, Institute for Data Processing and Electronics) for discussions on electronics and HV development, and Markus Steidl and Klaus Eitel (KIT, Institute for Astroparticle Physics) for their feedback on technical aspects and project management. We also thank David Wolf (University of Zurich) for support with the SiPM readout boards, and Bernd Berger (KIT, Electronics Workshop of the Institute of Experimental Particle Physics) for producing them. This work was supported by KIT's Young Investigator Group Preparation Program (YIG Prep Pro), the Helmholtz Initiative and Networking Fund (grants VH-NG-21-02 and W2/W3-118), and the KIT Center Elementary Particle and Astroparticle Physics (KCETA).

\bibliographystyle{JHEP}   
\bibliography{biblio}          

\end{document}